\documentclass[10pt, conference]{IEEEtran}
\IEEEoverridecommandlockouts
\usepackage{cite}
\usepackage{algorithmic}
\usepackage{graphicx}
\usepackage{textcomp}
\usepackage{xcolor}
\usepackage{enumitem}
\usepackage{amsmath}   
\usepackage{amssymb}
\usepackage{amsthm}            
{
	\theoremstyle{plain}
	\newtheorem{assumption}{Assumption}
}
\usepackage{multirow}

\usepackage{dblfloatfix}    
\usepackage[font=footnotesize]{subfig}

\newcommand{\hide}[1]{}

\newcommand{\assets}[0]{{k}}
\newcommand{\window}[0]{{w}}
\newcommand{\windowBackward}[0]{{b}}
\newcommand{\windowForward}[0]{{f}}

\newcommand{\sfigbigbig}[0]{7.5cm}
\newcommand{\sfigwhole}[0]{18.0cm}
\newcommand{\sfigbig}[0]{5.65cm}
\newcommand{\sfigmedium}[0]{4.8cm}

\newcommand{\sfigsmall}[0]{3.8cm}

\newcommand{\riskAnonym}{{\theta}}
\newcommand{\return}{{r}}

\newcommand{\priceSequence}{{M}}
\newcommand{\backwardSequence}{{\priceSequence_{\windowBackward}}}
\newcommand{\forwardSequence}{{\priceSequence_{\windowForward}}}

\newcommand{\diversification}{{x}}
\newcommand{\assetCounter}{{i}}
\newcommand{\simulationSet}{{S}}
\newcommand{\simulations}{{n}}
\newcommand{\price}{{p}}
\newcommand{\priceStart}{{s}}
\newcommand{\priceEnd}{{e}}
\newcommand{\portfolioReturn}{{\return_{p}}}
\newcommand{\expectedReturn}{{\mu}}
\newcommand{\portfolioRisk}{{\sigma_{p}^2}}
\newcommand{\expectedPortfolioReturn}{{\expectedReturn_{p}}}

\newcommand{\expectedReturnVector}{{\expectedReturn}}
\newcommand{\covarianceMatrix}{{\Sigma}}

\newcommand{\analysisValue}{{a}}
\newcommand{\analysisVector}{{A}}

\newcommand{\efficientSolutions}{{X}}
\newcommand{\riskLevel}{{\zeta}}
\newcommand{\totalRiskLevels}{{Z}}
\newcommand{\returnLevel}{{\hat{\return}}}

\newcommand{\sharpe}{\rho}
\newcommand{\riskFreeReturn}{{\overline{\return}}}

\newcommand{\generator}{{G}}

\newcommand{\simulatedSequence}{{\hat{\forwardSequence}}}
\newcommand{\syntheticData}{{\hat{\priceSequence}}}
\newcommand{\latent}[0]{\lambda}
\newcommand{\simTconvLayers}[0]{{l}}

\newcommand{\discriminator}{{D}}
\newcommand{\critic}{{c}}

\newcommand{\argmin}[1]{\underset{#1}{\operatorname{arg}\!\operatorname{min}}\;}

\newcommand{\R}{\mathbb{R}}

\begin{document}

\title{PAGAN: Portfolio Analysis with Generative Adversarial Networks}

\author{
\IEEEauthorblockN{
\emph{Giovanni Mariani$^{\star,\circ,1}$ \qquad Yada Zhu$^{\dagger}$ \qquad Jianbo Li$^{\ddagger}$ \qquad Florian Scheidegger$^{\star}$ }\\ 
\emph{Roxana Istrate$^{\star}$ \qquad Costas Bekas$^{\star}$ \qquad A. Cristiano I. Malossi$^{\star}$ }\\
\\
$^{\star}$ IBM Research -- Zurich, CH \qquad \qquad \qquad $^{\dagger}$ IBM Research, US \\
    $^{\ddagger}$ Three Bridges Capital, US  \qquad \qquad \qquad $^{\circ}$ Qualcomm, US
}
}

\maketitle

\begin{abstract}
	Since decades, the data science community tries to propose prediction models of financial time series. Yet, driven by the rapid development of information technology and machine intelligence, the velocity of today's information leads to high market efficiency. Sound financial theories demonstrate that in an efficient marketplace all information available  today, including expectations on future events, are represented in today prices whereas future price trend is driven by the uncertainty. This jeopardizes the efforts put in designing prediction models. To deal with the unpredictability of financial systems, today's portfolio management is largely based on the Markowitz framework which puts more emphasis in the analysis of the market uncertainty and less in the price prediction. The limitation of the Markowitz framework stands in taking very strong ideal assumptions about future returns probability distribution.
	
	To address this situation we propose PAGAN, a pioneering methodology based on deep generative models. The goal is modeling the market uncertainty that ultimately is the main factor driving future trends. The generative model learns the joint probability distribution of price trends for a set of financial assets to match the probability distribution of the real market. Once the model is trained, a portfolio is optimized by deciding the best diversification to minimize the risk and maximize the expected returns observed over the execution of several simulations. Applying the model for analyzing possible futures, is as simple as executing a Monte Carlo simulation, a technique very familiar to finance experts. The experimental results on different portfolios representing different geopolitical areas and industrial segments constructed using real-world public data sets demonstrate promising results\footnote{Giovanni Mariani contributed to this work while employed at IBM. It was published after he joined Qualcomm Netherlands B.V..}$^,$\footnote{IBM, the IBM logo, and ibm.com are trademarks or registered trademarks of International Business Machines Corporation in the United States, other countries, or both. Other product and service names might be trademarks of IBM or other companies.}.
\end{abstract}


\section{Introduction}
Financial markets play an important role on the economical and social organization of modern society. Trigger by recent machine intelligence, investment industry is experiencing a revolution. Despite this, nowadays financial portfolio management is still largely based on linear models and the Markowitz framework  \cite{Papenbrock2016}, known as Modern Portfolio Theory (MPT). The MPT aims to achieve portfolio diversification while minimizing specific risks and determining the risk-return trade-offs for each asset \cite{Markowitz1959}. Despite the significant and long lasting impact,  MPT has been criticized for its ideal assumption on the financial system and data. The MPT heavily relies on the accurate estimate of the future returns and volatilities for each asset and their correlations. However, market price forecast is one of the main challenges in the time series literature due to its highly noisy, stochastic and chaotic nature \cite{Tsay2010}. More importantly,  the velocity of today's information systems enables traders and investors to take decisions based on real-time market updates. This fact leads to high market efficiency, i.e. the price agreed between the different parties involved in the trade accounts for all available information about the present as well as all the forecasts about the future that are inferable today. Under high efficient market, as soon as a trusted long-term forecast is provided, this forecast would be consumed by traders in the short term and have a direct impact already on current price whereas future price variations would again be uncertain \cite{timmermann2004}. With the continuous advancement of the financial transactions and the information systems, the market becomes increasingly efficient, so does the challenge of forecasting market price.  

To address the challenges presented in portfolio management, in this paper, we present a pioneering study about portfolio analysis with generative adversarial networks (PAGAN). PAGAN directly models the market uncertainty, the main factor driving future price trend, in its complex multidimensional form, such that the non-linear interactions between different assets can be embedded. We propose an optimization methodology that utilizes the probability distribution of real market trends learnt from training PAGAN to determine the best portfolio diversification minimizing the risk and maximizing the expected returns observed over the execution of multiple simulations.

The main contributions of the paper are summarized below.

\begin{itemize}[ topsep=5pt]
	\item \textbf{Problem Setting.} Different from existing work and practice, in this paper, we aim to  model the market uncertainty conditioned with the most recent past, allowing automatically learning nonlinear market behavior and non-linear dependencies between different financial assets and generating many realistic future trends given today situation;
	\item \textbf{Optimization methodology.} We propose a generative model on real time series named PAGAN and an optimization methodology to use the PAGAN model for solving the portfolio diversification problem. The probability distribution learnt by PAGAN enables us to play with different diversification options to trade off risk for expected returns. The final diversification to be implemented is the one realizing the sweet point on the predicted risk-return efficient frontier, given a target user-selected risk level;
	\item \textbf{Experiments on Real Data Sets.} We evaluate the proposed methodology on two different portfolio representative of different markets and industrial segments. Results demonstrate that the proposed approach is able to realize the risk-return trade off and significantly outperform the traditional  MPT.
\end{itemize}

The rest of the paper is organized as follows. After introducing the MPT in Section 2, we brief review of the related work in Section 3. We introduce the proposed PAGAN methodology in Section 4, including market uncertainty modeling, the generative network architecture, as well as the optimization approach for determining portfolio diversification. Section 5 provides some experimental results demonstrating the effective of the PAGAN on real-world finance assets and data. Finally, we conclude the paper in Section 6.

\section{The Markowitz's Framework}
\label{sec:motivation}


Let us consider a fully invested long only portfolio consisting of a set of financial assets and define a portfolio diversification strategy as a vector $\diversification$ where $\diversification_{i}$
is the amount of capital we invest in the $i$th asset. 
We optimize the portfolio diversification $\diversification$
to maximize the expected portfolio returns and minimize the portfolio risk by taking an educated guess on the probability distribution of future assets' returns. According to Markowitz's framework \cite{Markowitz1959}, the educated guess on the probability distribution of assets' returns $\return$ is given by the following assumption.
\begin{assumption}\label{as:mark}
	$\return \sim \mathcal{N}(\expectedReturnVector, \covarianceMatrix)$ where $\return$ is the returns vector ($\return_i$ is the
	return for asset $i$), 
	$\expectedReturnVector$ is the expected mean returns vector, and $\covarianceMatrix$ is the covariance matrix.
	The expected mean returns $\expectedReturnVector$, and the returns covariance matrix $\covarianceMatrix$ are
	estimated from the past observations and assumed constant in the future.
\end{assumption}

Given a portfolio diversification $\diversification$, the portfolio future returns $\portfolioReturn(\diversification)$ are
originated by the linear combination of the returns of individual assets. Given Assumption \ref{as:mark}:
\begin{eqnarray}
\portfolioReturn(\diversification) \sim \mathcal{N}\big(\expectedPortfolioReturn(\diversification), ~ \portfolioRisk(\diversification)\big)\\
\label{eq:retDef}\expectedPortfolioReturn(\diversification) = \diversification \cdot  \expectedReturn_i\\
\label{eq:riskDef}\portfolioRisk(\diversification) = \sum_{i}\sum_{j} \covarianceMatrix_{i,j} * \diversification_{i} * \diversification_{j}\\
\label{eq:constraint0}\sum_i \diversification_i = 1 ~ ~ \bigwedge ~ ~ \forall_i ~ \diversification_i \geq 0
\end{eqnarray}

\begin{figure}[t!]
	\centering
	\resizebox{\sfigsmall}{!}{\includegraphics[trim={0, 0, 0, 0}, clip]{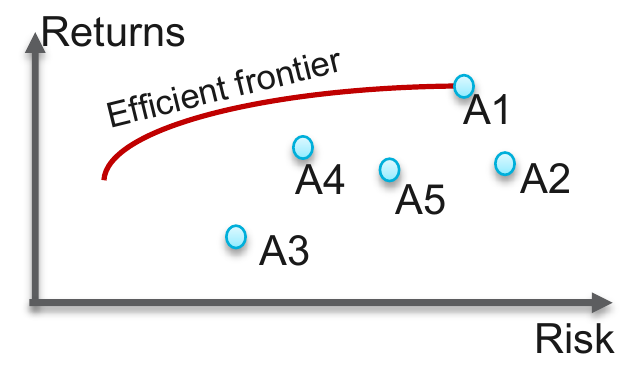}}
	\caption{{Example of efficient frontier delineating the best risk-return trade off achievable with Markowitz approach.  $\{A_1 ... A_5\}$ indicate individual assets.}}
	\label{fig:efficientFrontier}
\end{figure}

\begin{figure}[t!]
	\centering
	\resizebox{\sfigbig}{!}{\includegraphics[trim={0, 0, 0, 0}, clip]{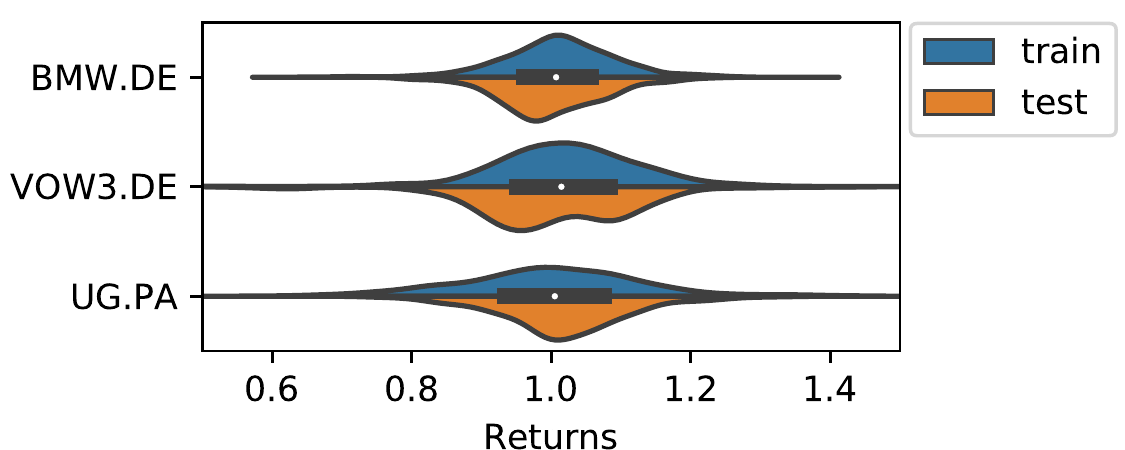}}
	\caption{{Probability distribution of monthly returns for three assets ($y$ axis, more information in Table \ref{tab:portfolios}) during two time periods:  \textit{a)} \textit{train} (up to the end of 2014),
			and \textit{b)} \textit{test} (from 2015 onwards). 
			Returns probability distribution differs for the two periods, thus
			Assumption \ref{as:mark} does not hold.
	 }}
	\label{fig:motivation}
\end{figure}

Equation \ref{eq:retDef} defines the portfolio returns expectation $\expectedPortfolioReturn$, to be maximized by selecting $\diversification$.
Equation \ref{eq:riskDef} defines the portfolio risk factor $\portfolioRisk$ to be minimized. Equation \ref{eq:constraint0} 
defines the optimization constraints.
This optimization problem can be solved
in closed form \cite{markowitz}. Figure \ref{fig:efficientFrontier} shows with a continuous line
an example of efficient frontier that one can identify in the risk--returns objective space
by diversifying the portfolio with a set of assets $\{A_1 ... A_5\}$. Every point in the line corresponds to a different diversification $\diversification$.
To reach the efficient frontier and minimize the risk,
it is necessary to leverage the covariance matrix $\covarianceMatrix$ and to include in the portfolio less correlated or
inversely-correlated assets (Equation \ref{eq:riskDef}).
Whereas Markowitz's identification of the efficient frontier has a very solid mathematical background, Assumption \ref{as:mark}
hardly applies in the real world for the following reasons: \textit{a)}
returns for an individual asset are not normally distributed,
\textit{b)} interactions between different assets may be non-linear whereas the covariance $\covarianceMatrix$ captures only linear dependencies, and
\textit{c)} future probability distribution of assets' returns may deviate from the past (Figure \ref{fig:motivation}).

In this work, we use generative models to solve all these three issues in one go with the proposed PAGAN model as follows:
\textit{a)} PAGAN does not rely on any preliminary assumption about the probability distribution of individual asset returns,
\textit{b)} the non-linearities of the neural network
implicitly embed non-linear interactions between different assets, and \textit{c)} PAGAN is explicitly designed
to take input as the current market situation and to model the future probability distribution of returns.

\section{Related Work}
\subsection{Portfolio Management}

To address the limitation of MPT, a main breakthrough is the introduction of conditional volatility models that allow returns' volatility to vary through time \cite{Bollerslev1988}.
These sophisticated statistical models assumes that relationships shift will eventually return to normal, and therefore tend to fail when the shifts in returns or correlations are more permanent \cite{Rapach2008}.  Recently, the application of machine learning to finance has drawn interests from both investors and researchers. However, the current work has been focused on adapting reinforcement learning (RL) frameworks to trading strategy \cite{Deng2017, Jiang2017, Almahdi2017}. To be specific, the market action of the RL agent defines portfolio weights. An asset with an increased target weight will be bought in with additional amount, and that with decreased weight will be sold. These works focus on intra-day decision, while we tackle the challenge of medium-long term portfolio diversification. In addition, all these RL algorithms assume a single agent and ideal trading environment, i.e., each trade is carried out immediately and has no influence on the market. In fact, the trading environment is essentially a multiplayer game with thousands of agents acting simultaneously and impact the market in a complicated way. Considering the complexity of financial market, combining machine learning with portfolio management remains relatively unexplored.

\subsection{GANs on Time Series}
Recently generative adversarial networks (GAN) has become an active research area in learning generative models. GAN is introduced by Goodfellow et al. \cite{Goodfellow2014}, where images patches are generated from random noise using two networks training simultaneously. The discriminative net $D$ learns to distinguish whether a given data instance is real or not, and a generative net $G$ learns to confuse $D$ by generating high quality data. Powered by the learning capabilities of deep neural networks, GANs have gained remarkable success in computer vision and natural language processing. \hide{\cite{Saatci2017}} 

However, to date there has been limited work in adopting the GAN framework for continuous time series data. One of the preliminary works in the literature produces polyphonic music with recurrent neural networks as both generator and discriminator \cite{Mogren2016}, and the other one uses conditional version of recurrent GAN to generate real-valued medical time series \cite{Hyland2017}. In these methods, the multiple sequences are treated as i.i.d. and fed to a uniform GAN framework. Stock price is largely driven by the fundamental performance of an individual company and the dynamic interactions of different stocks are embedded in their prices. Thus the i.i.d. assumption of multiple sequences is too restricted for portfolio analysis. Li et al. \cite{Li2018} adopt Long Short Term-Recurrent Neural Networks (LSTM-RNN) in both the generator and discriminator of GAN to capture the temporal dependency of time series. Zhou et al. \cite{Zhou2018} apply the superposition of an adversarial loss and a prediction loss to improve a traditional LSTM-based forecast network. Yet, in Zhou there is no actual generative model in the sense that, once trained, the prediction model of Zhou returns a single deterministic forecast given an input price sequence whereas in the proposed PAGAN model we provide the capability of simulating possible future situations by sampling future prices from a posteriori probability distribution learnt with the adversarial training process.

\subsection{Stock Market Forecast}
Stock market forecast is one of the most challenging issues among time series forecasting \cite{Tsay2010} due to chaotic dynamics of the markets. Traditionally, statistical methods, such as autoregressive model, moving average model, and their combinations, are widely used.  However, these methods rely on restricted assumptions with respect to the noise terms and loss functions. During the past decades, machine learning models, such as artificial neural networks \cite{Kara2011, Ghiassi2013} and support vector regression \cite{WeiHuang2005, Sheta2015} have been applied to predict future stock prices and price movement direction. More recently,  a deep convolutional neural network is applied to predict the influence of events on stock movements \cite{Ding2015}. Kuremoto et al. \cite{Kuremoto2014} present a deep belief network with restricted Boltzmann machines and Bao et al. \cite{Bao2017} investigate  autoencoders and LSTM for short-term stock price forecast. However, with the continuous advancement of the financial transactions and the information systems, financial market becomes increasingly efficient.  This leads to increased market uncertainty and challenge of forecasting market price. Instead of predicting market trend, we for the first time learn modeling the uncertainty of the marketplace in its sophisticated multidimensional form for portfolio management.  


\section{Proposed methodology}
\subsection{Overview}
\begin{figure}[b!]
\centering
\resizebox{\sfigsmall}{!}{\includegraphics {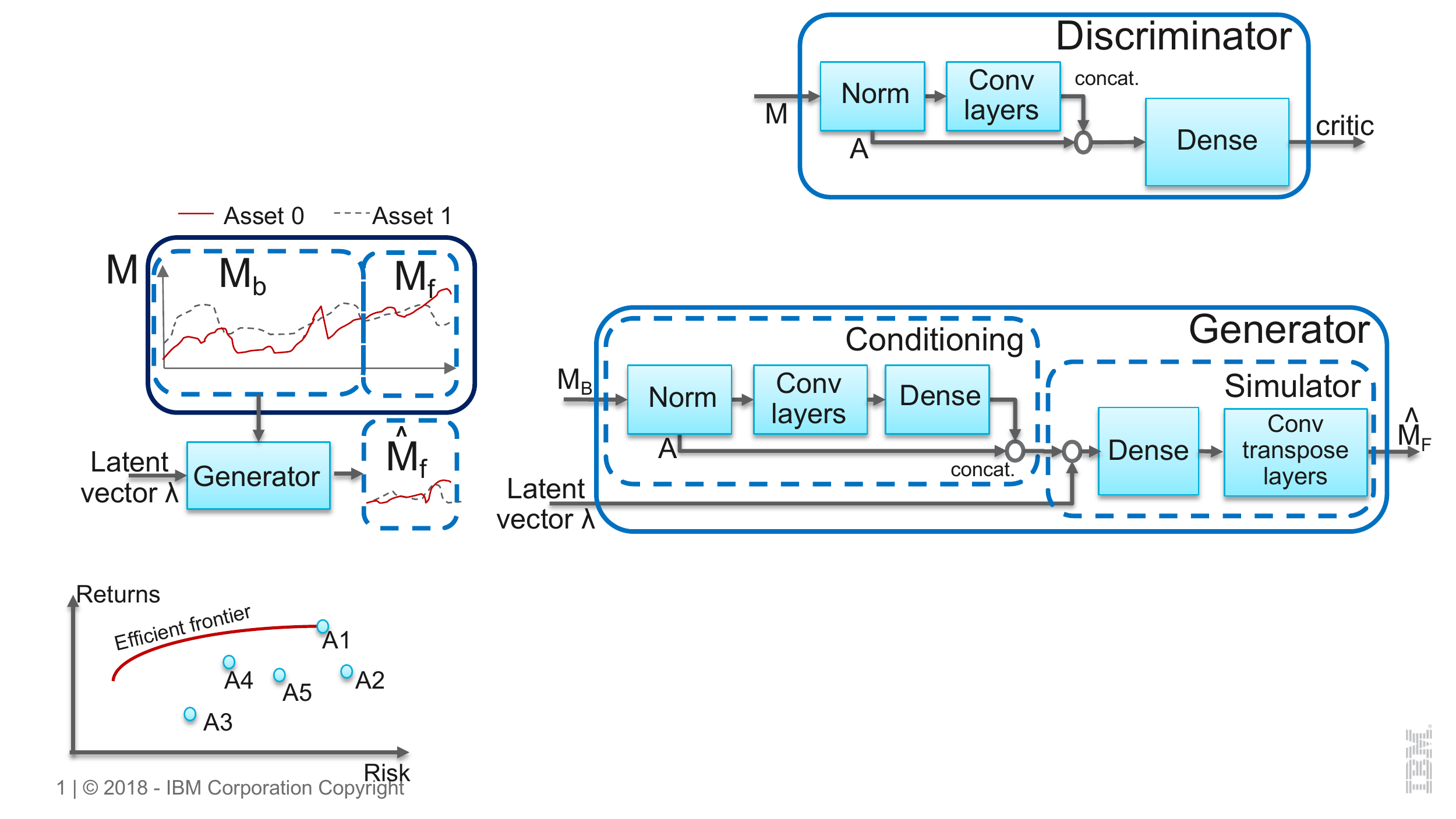}}
\caption{{Conceptual overview of the input and output of the proposed PAGAN generator. }}
\label{fig:overview}
\end{figure}
In this work we deal with deep-learning networks for time series data. As observed by other authors \cite{shaojie2018},
1D convolutional networks are an effective tool to process time series and can outperform traditional recurrent networks
in terms of both result quality and performance.
For this reason we process time series by means of convolutional networks and represent the
asset-price trends
as a matrix $\priceSequence$ with
$\assets$ rows (financial assets) and $\window$ columns (days), $\priceSequence\in \R^{\assets\times\window}$.
The deep-learning networks process the time information by convolving along the time dimension.

Our aim in this work is to model the probability distribution of the asset-price trends
for the future $\windowForward$ days given the current market situation represented by the latest observed
$\windowBackward$ days. We consider the matrix $\priceSequence$ to span the whole analysis length:
$\window=\windowForward + \windowBackward$. Thus, $\priceSequence$ is composed of two parts: \textit{a)}
the known past $\backwardSequence$ of length $\windowBackward$, and \textit{b)} the unknown future $\forwardSequence$ of length $\windowForward$.
We apply a generative deep-neural network $\generator$ to learn
the probability distribution of future price trends $\forwardSequence$ within the target future horizon $\windowForward$
given the known recent past $\backwardSequence$, and a prior distribution of a random latent vector.
Figure \ref{fig:overview} shows a graphical interpretation
of what the matrix $\priceSequence$ represents and the input and output of
the generator $\generator$.
Formally the generative model returns
a synthetic possible future matrix $\simulatedSequence$ (a simulation) as a function:

\begin{equation}
    \simulatedSequence = \generator(\backwardSequence, \latent),
\end{equation}
where $\latent$ is the latent vector sampled from
a prior distribution.
In practice,
$\latent$ represents the unknown future events and phenomena impacting the marketplace. The known past $\backwardSequence$
is used to condition the probability distribution of the future $\simulatedSequence$ based on the most updated market situation.
The generator $\generator$ is a generative network which weights are learnt to let $\simulatedSequence$ match the probability distribution
of $\forwardSequence$ given the past $\backwardSequence$ on a set of training data. The generator $\generator$
is trained in adversarial mode against a discriminator $\discriminator$ with the goal of minimizing the Wasserstein distance between synthetic data $\simulatedSequence$
and real data $\forwardSequence$, based on historical observations.
The training process has the goal
to approximate the real posterior probability distribution
$P(\forwardSequence | \backwardSequence)$
with the surrogate probability distribution $P(\simulatedSequence | \backwardSequence)$,
given the prior distribution of $\latent$. In this work we use the normal distribution for the prior $\lambda$.

To implement the adversarial training process we consider a discriminator network
$\discriminator$ to take as input the overall price matrix $\priceSequence$,
that is the concatenation of the conditioning $\backwardSequence$ and either the synthetic data $\simulatedSequence$ or the real data $\forwardSequence$.
The discriminator output is a critic value $\critic=\discriminator(\priceSequence)$.
The discriminator is trained to minimize $\critic$
for the real data and maximize it for synthetic data, whereas the generator $\generator$
training goal is to minimize $\critic$ for the synthetic data.
In this work we apply WGAN-GP methodology \cite{gulrajani2017}.

\subsection{Deep-learning architecture}
\begin{figure}[t!]
	\centering
	\subfloat[{Generator.}]{
		\label{fig:genArch}
		\resizebox{\sfigbigbig}{!}{\includegraphics{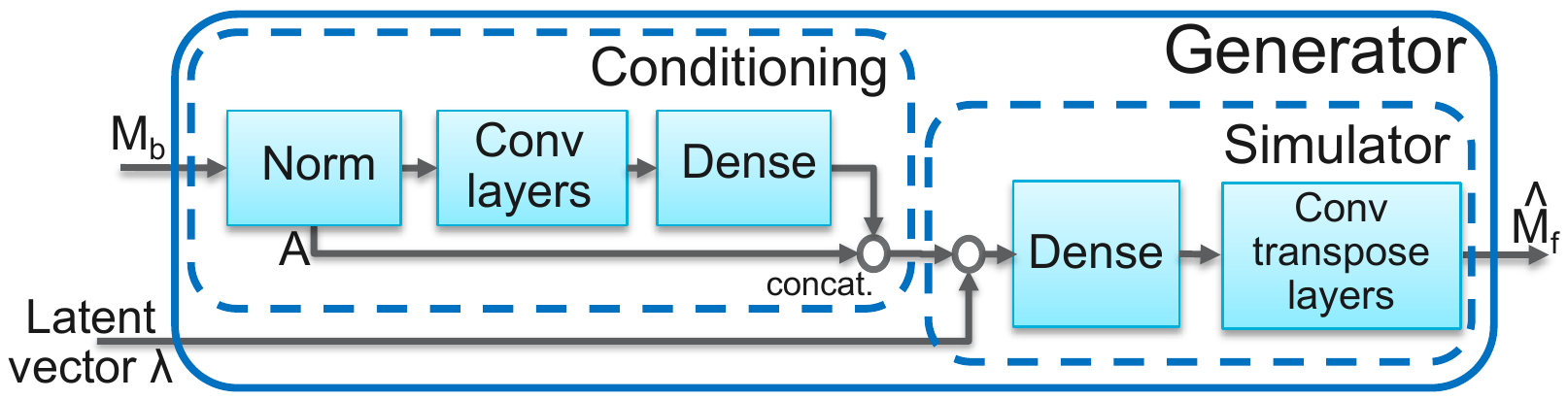}}
	}
	\qquad
	\subfloat[{Discriminator.}]{
		\label{fig:disArch}
		\resizebox{\sfigmedium}{!}{\includegraphics{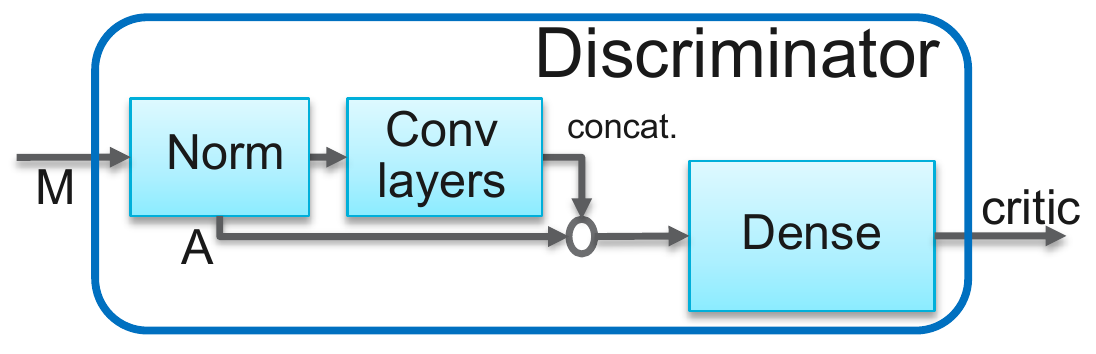}}
	}
	\caption{
		Architectures of the PAGAN generative and discriminative models.
	}
	\label{fig:architecture}
\end{figure}

\textbf{Data normalization.} We consider the \textit{adjusted close} price $\price$ for each financial asset.
During training, given a time window of $\window=\windowBackward+\windowForward$ days, we normalize the prices $\price$ for each asset
to fit in the range $[-1, 1]$ for the initial $\windowBackward$ days.
The normalization output is the daily asset price variation $\price(t) - \price(t-1)$ computed in this normalized scale.
Whereas
normalizing to the range $[-1, 1]$ enables us to expose to the neural networks
values limited within a reasonable range, the normalization removes from the data information about the price-variability
within the given window $\window$.
Since the normalized values of $\backwardSequence$ always range between $[-1, 1]$, the presence of long tails
in the data is normalized out. To feed into the neural network information about the observed price variability and possible presence of
long-tails, during the normalization procedure we also compute an \textit{analysis value}
$\analysisValue$ for each asset:
\begin{equation}\label{eq:analysisValue}
\analysisValue = \frac{\price_{max} - \price_{min}}{\price_{mean}},
\end{equation}
where $\price_{max}$, $\price_{min}$, and $\price_{mean}$ are respectively the maximum, minimum and mean values of the price $\price$
in $\backwardSequence$ for a given asset. Let's define the analysis vector $\analysisVector$ as the vector representation of
$\analysisValue$ to consider multiple assets.
Figure \ref{fig:architecture} shows the architecture of PAGAN generator and discriminator and explicitly clarify the presence of
this non-traditional normalization process (\textit{Norm}). 


\textbf{Generator.} 
The generator $\generator$ takes as input the price sequence $\backwardSequence$
and the latent vector $\latent$ (Figure \ref{fig:genArch}). $\generator$ is composed of two consecutive parts: \textit{a)}
a \textit{conditioning} network to compute an inner representation of the past price sequence $\backwardSequence$, and \textit{b)} a \textit{simulator}
network to generate the simulation of future price trends.

The \textit{conditioning} input is the most recent market trend $\backwardSequence$. After the normalization, we
apply a set of 1D convolution and dense layers as described in details in Appendix \ref{sec:hyperparams}. 
The \textit{conditioning} output depends only on
$\backwardSequence$ and is used to condition the probability distribution of the synthetic data
$P\big(\simulatedSequence ~ |~ \backwardSequence\big)$, Equation \ref{eq:Pgenerator}.

The \textit{simulator} network takes as input the \textit{conditioning} output and the latent vector $\latent$,
and generates as output a simulation of future market prices $\simulatedSequence$.
The \textit{conditioning}, and the \textit{simulator} together implement the generator $\generator$ and are trained at once
against the discriminator $\discriminator$ with the traditional adversarial approach.

\textbf{Discriminator.} The discriminator takes as input either \textit{a)} the real data $\priceSequence$, concatenation of $\backwardSequence$ and $\forwardSequence$,
or \textit{b)} the synthetic data $\syntheticData$, concatenation of $\backwardSequence$ and $\simulatedSequence$. 
The discriminator output is computed by the network shown in Figure \ref{fig:disArch}.

\textbf{Architectural parameters.} Appendix \ref{sec:hyperparams} in the supplementary material
describes all parameter and other details of the generator and discriminator networks.

\subsection{Portfolio optimization}

Once the training process is completed, the generator $\generator$ is able
to synthesize realistic future trends
$\simulatedSequence = \generator(\backwardSequence, \latent)$.
We use these synthetic simulations to numerically estimate
the expected risks and returns
for different portfolio diversification options $\diversification$.
We thus execute a portfolio optimization
on the estimated posterior probability distribution:
\begin{equation}
P\big(\simulatedSequence ~ |~ \backwardSequence\big), ~ ~ \simulatedSequence=\generator(\backwardSequence, \latent)
\end{equation}
given the known prior distribution of $\latent$ and the conditioning $\backwardSequence$.

For a given conditioning $\backwardSequence$,
let us consider a set $\simulationSet$
of $\simulations$ simulations $\simulatedSequence\in\simulationSet$ sampled from
$P\big(\simulatedSequence ~ |~ \backwardSequence\big)$ by evaluating the generative model
$\generator(\backwardSequence, \latent)$ on
different extractions of $\latent$.
Let us define also the return vector
function $\return(\simulatedSequence)$ where the $\assetCounter$th element
$\return_{\assetCounter}$ is the return obtained by the $\assetCounter$th asset at the end of the simulation horizon
for one simulation $\simulatedSequence$ given by ${\return}_{\assetCounter}=\priceEnd_{\assetCounter}/\priceStart_{\assetCounter}-1$, where $\priceEnd$ is the asset price at the end of the simulation $\simulatedSequence$,
and $\priceStart$ is the price at the beginning of the simulation. Since the constant in the definition of  ${\return}_{\assetCounter}$ does not impact the optimization results, in the work, we use
\begin{equation} \label{eq:returnDefinition}
{\return}_{\assetCounter}=\priceEnd_{\assetCounter}/\priceStart_{\assetCounter}
\end{equation}
The portfolio returns achieved with the diversification $\diversification$
for a given simulation $\simulatedSequence$ is:
\begin{equation}\label{eq:Pgenerator}
    \portfolioReturn(\diversification, \simulatedSequence)=\diversification\cdot\return(\simulatedSequence)
\end{equation}

The simulations $\simulatedSequence\in\simulationSet$ sampled from the probability distribution
$P\big(\simulatedSequence ~ |~ \backwardSequence\big)$ are used to
infer the probability distribution
$P\big(\portfolioReturn(\diversification, \simulatedSequence) ~ |~ \backwardSequence, \diversification\big)$.

The portfolio optimization problem is defined as in the traditional Markowitz' optimization approach (Section \ref{sec:motivation}),
yet it is executed on the predicted future probability distribution $P\big(\portfolioReturn(\diversification, \simulatedSequence) ~ |~ \backwardSequence, \diversification\big)$
that is non-normal and includes non-linear interactions between the different assets.
For instance, the optimization goal is to identify the configurations of $\diversification$
that maximize the expected returns
$\expectedPortfolioReturn=\mathbb{E}\big(\portfolioReturn(\diversification, \simulatedSequence) ~ |~ \backwardSequence, \diversification\big)$
and minimize a risk function $\riskAnonym(\backwardSequence, \diversification)$.
Both $\riskAnonym$ and $\expectedPortfolioReturn$ are estimated on the
base of the simulation samples $\simulatedSequence\in\simulationSet$.
In this framework, the risk function $\riskAnonym(\diversification)$ can be any metric such as the \textit{value at risk} \cite{chen2013}, or the \textit{volatility}.
Without loss of generality, we use
the estimated volatility (variance) that enables us to evaluate the approach directly with respect to the traditional Markowitz's methodology.
The optimization problem is thus formalized as:
\begin{eqnarray}
\label{eq:obj0}\max_\diversification \expectedPortfolioReturn(\diversification ~ | ~ \simulationSet), \\
\label{eq:obj1}\min_\diversification \riskAnonym(\diversification ~ | ~ \simulationSet), \\
\label{eq:retDefPAGAN}\expectedPortfolioReturn(\diversification)=\mathbb{E}\big(\portfolioReturn(\diversification, \simulatedSequence) ~ | ~ \backwardSequence, \diversification\big), \\
\label{eq:riskDefPAGAN}\riskAnonym(\diversification)=\portfolioRisk(\diversification)={\mathbb{V}ar\big(\portfolioReturn(\diversification, \simulatedSequence) ~ |~ \backwardSequence, \diversification\big)}
\end{eqnarray}
where Equations \ref{eq:obj0}, and \ref{eq:obj1} are the target objectives.
We solve the optimization problem by means of a multi-objective genetic algorithm,
the NSGA-II \cite{deb2002} to provide a trade-off between expected returns and risk.
The output of the NSGA-II is a set $\efficientSolutions(\simulationSet)$ that depends on the simulations $\simulationSet$.
Elements $\diversification\in\efficientSolutions$  are Pareto-optimal diversifications trading off returns and risk.
The decision of what diversification strategy $\diversification\in\efficientSolutions$ to use
is left to the end user depending on its own goals.



\section{Experimental results}
\label{sec:res}

\subsection{Portfolios setup}

We apply public available data Yahoo Finance \cite{yahooFinance} to back-testing the proposed PAGAN approach. A matrix $\priceSequence$ representing the price trend for a window of length $\window$, i.e.  2006-05-01  to 2015-01-01, is considered train data, and 2015-01-012 to 2018-06-30 is considered as test data. The proposed PAGAN methodology learns the generative model $\generator$ from train data and applies it for optimizing portfolio for the test data.

In this work, we investigate two different portfolios representing different geopolitical areas and industrial segments. 
To select the assets to be included in the portfolios, we mainly looked
at the following three criteria. \textit{a) Data availability}: we only include assets for which data is available from at least 2006, given Yahoo Finance data source.
\textit{b) Currency homogeneity}: whereas we consider different portfolios
with different currencies, in a single portfolio we include assets traded in a single currency. This is  is not an actual limitation, yet it facilitates our evaluation process.
\textit{c) Data correctness}: we identified some erroneous data from Yahoo Finance, e.g. \textit{NaN} values or fluctuation of $10\times$ in asset price
lasting a single day, etc.. Whereas these errors are rare, we systematically discard the associated assets.

In the considered portfolios we include a set of lower-risk securities (e.g. the overall market index).
This is a common approach in portfolio optimization and it enables us a wide range of options
to trade off between low-risk securities and high-returns ones. The considered portfolios are detailed
in Table \ref{tab:portfolios} and summarized as follows:
\begin{itemize}
	\item \textit{US general market} (\textit{usgen}): a set of US firms from different market segments, such as IT (GOOG, MSFT), healthcare (CELG, PFE),
	 energy (HES, XOM), and consumer staples (KR, WBA). We also include three ETFs tracking the overall US markets (SHY, IYR, IYY).
	\item \textit{EU automotive} (\textit{eucar}): four of the most well known European automotive companies (BMW.DE, FCA.MI, UG.PA, VOW3.DE).
	We balance these shares with two EU indices tracking the German and French stock markets (~$\hat{ }$~FCHI, ~$\hat{ }$~GDAXI).
\end{itemize}

\begin{table}[t!]
	\centering
	\scriptsize
	\caption{List of assets in the portfolios \textit{usht}, and \textit{eucar}.} 
	\setlength\tabcolsep{0.15cm}
	\begin{tabular}{|c||   c|c|c|c|c| }
		\hline
		& \textbf{Ticker} & \textbf{Type}  & \textbf{Industry} & \textbf{Description} & \textbf{Cur.} \\ 
		\hline
		\hline
		\multirow{11}{*}{\textit{usgen}}	& GOOG	&	Share	&	IT	&	Alphabet	&	USD	\\
		& MSFT	&	Share	&	IT	&	Microsoft	&	USD	\\
		\cline{2-6}
		& \textit{CELG}	& Share & Healthcare & Celgene & USD \\
		& PFE	&	Share	&	Healthcare	&	Pfizer	&	USD	\\
		\cline{2-6}
		& HES	&	Share	&	Energy	&	Hess	&	USD	\\
		& XOM	&	Share	&	Energy	&	Exxon Mobil	&	USD	\\
		\cline{2-6}
		& KR	&	Share	&	Consumer staples	&	The Kroger	&	USD	\\
		& WBA	&	Share	&	Consumer staples	&	Walgreens Boots Alliance	&	USD	\\
		\cline{2-6}
		& IYY	&	ETF	&	Dow Jones	&	iShares Dow Jones	&	USD	\\
		& IYR	&	ETF	&	Real estate	&	iShares US Real Estate	&	USD	\\
		& SHY	&	ETF	&	US treasury bond	&	iShares Treasury Bond &	USD	\\
		\hline
		\hline
		\multirow{6}{*}{\textit{eucar}}	&	BMW.DE	&	Share	&	Automotive	&	BMW	&	EUR	\\
		& 	FCA.MI	& Share & Automotive &	Fiat Chrysler Automobiles	& EUR \\
		& UG.PA	&	Share	&	Automotive	&	Peugeot &	EUR	\\
		& VOW3.DE	&	Share	&	Automotive	&	Volkswagen	&	EUR	\\
		\cline{2-6}
		& $\hat{ }$~FCHI	&	Index	&	French market	&	CAC 40	&	EUR	\\
		& $\hat{ }$~GDAXI	&	Index	&	German market	&	DAX	&	EUR	\\
		\hline
	\end{tabular}
	\label{tab:portfolios}
\end{table}

\begin{figure*}[t!]
	\centering
	\resizebox{\sfigwhole}{!}{\includegraphics{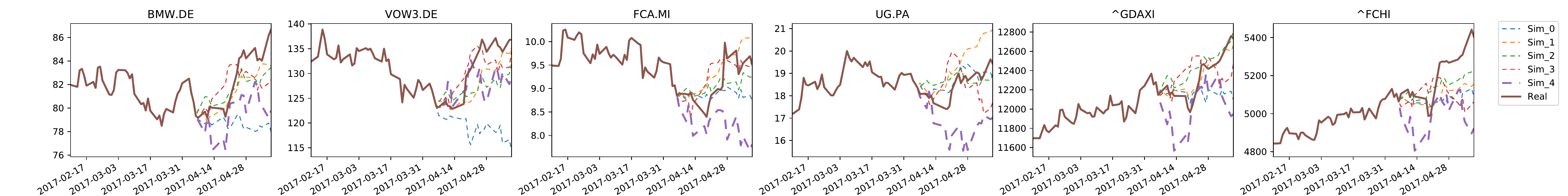}}
	\caption{{Actual price trend (solid line) for the \textit{eucar} portfolio between 2017-02 and 2017-04,
			and five representative simulations generated by PAGAN (dashed lines).
			Observe that simulations are correlated along the different assets, e.g. \textit{Sim\_4} represent the possibility of a general loss in the considered markets.
	}}
	\label{fig:resSimulations}
\end{figure*}

\subsection{Benchmarks}

\textbf{Markowitz.} We benchmark the proposed PAGAN approach with respect to the Markowitz' modern portfolio optimization \cite{markowitz}.
Since the genetic algorithm in PAGAN returns a discrete set of optimal diversifications $\diversification\in\efficientSolutions$
whereas the Markowitz methodology solves the optimization problem in a continuous form, to apply a
simple and fair comparison
we define a set of discrete risk levels $\riskLevel\in\{1, .., \totalRiskLevels\}$, with $\totalRiskLevels$
 an arbitrary Integer. In this work, $\totalRiskLevels=25$.
We define the target return $\returnLevel(\riskLevel)$ for the risk level $\riskLevel$ as follows.
The lowest risk level has a return goal of $\returnLevel(1)=1$.
Given Equation \ref{eq:returnDefinition}, this goal defensively
aims not to lose. We set an relatively large value for the maximum return level $\returnLevel(\totalRiskLevels)$. In this work
we use $\returnLevel(Z)=2\return_{max} -1$, where $\return_{max}$ is the maximum returns observed for any asset in the portfolio along the training period.
The factor 2 enables us to search for higher returns than those observed during training if these are considered possible by the generator $\generator$.
The constant $-1$ is introduced to compensate the fact that $\return=1$ is a non-loosing nor-winning policy (Equation \ref{eq:returnDefinition}).
Target returns for other risk levels $\returnLevel(\riskLevel)$ are uniformly spread:
\begin{equation}\label{eq:riskDefinition}
\returnLevel(\riskLevel)=\returnLevel(1) + (\riskLevel-1) \times \frac{\returnLevel(\totalRiskLevels) - \returnLevel(1)}{\totalRiskLevels-1}
\end{equation}
Once the efficient solutions $\efficientSolutions$ are found by the genetic algorithm, we select an optimal diversification $\diversification_{\riskLevel}$ for each
risk level $\riskLevel$ as:
\begin{equation}
\diversification_{\riskLevel} = \argmin{\diversification\in\efficientSolutions}|\returnLevel(\totalRiskLevels) - \return(\diversification)|
\end{equation}

\textbf{Default.} Given the continuously changing market situations, Assumption \ref{as:mark} used by the Markowitz methodology
hardly applies in general. Yet, there are limited other well-established approaches to predict the probability distribution of future market returns,
that makes the Markowitz's approach a widely-used and well-accepted method. The proposed PAGAN methodology is explicitly meant
to cope with this lack of sound alternative methodologies.
Though, in the recent years, markets have performed very differently from the past because of ending of QE, geopolitical situations,
trade wars, and scandals such as the Volkswagen emission scandal in September 2015. The return probability distribution differs
significantly between the \textit{train} and \textit{test} periods. For this reason we include as second benchmark (\textit{default})
a simple random optimization to verify how well one could perform by investing in randomly-selected assets.
Every day we sample $\totalRiskLevels$ diversifications $\diversification$ at random. We assign a risk level $\riskLevel$
to each of these diversification by sorting and enumerating them accordingly
to their expected returns as defined in Equation \ref{eq:retDef}.

\begin{figure*}[t!]
	\centering
	\subfloat[{\textit{usgen}, 5-days horizon.}]{
		\resizebox{\sfigbig}{!}{\includegraphics{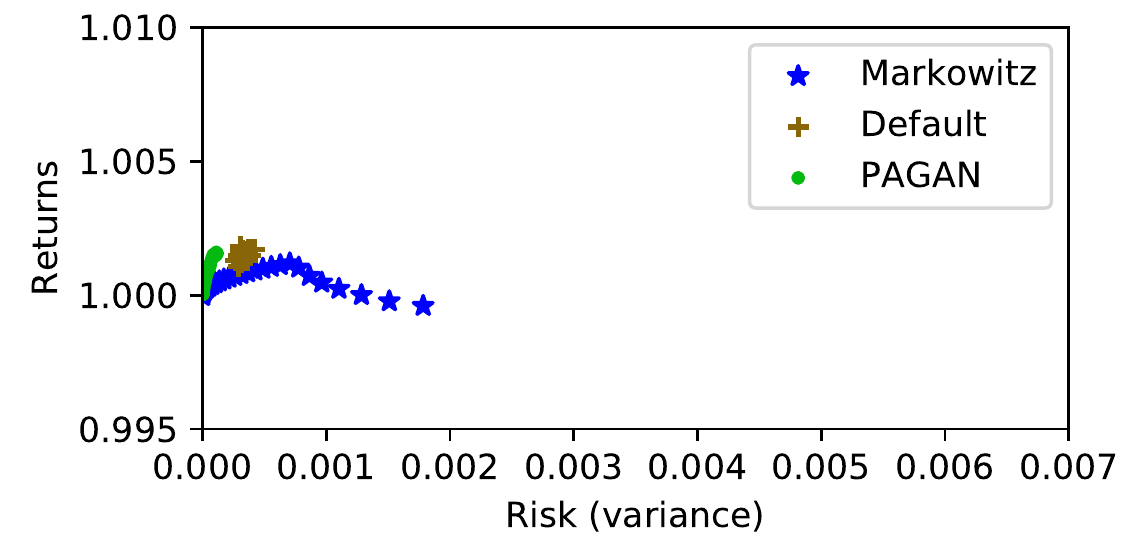}}
	}
	\subfloat[{\textit{usgen}, 10-days horizon.}]{
		\resizebox{\sfigbig}{!}{\includegraphics{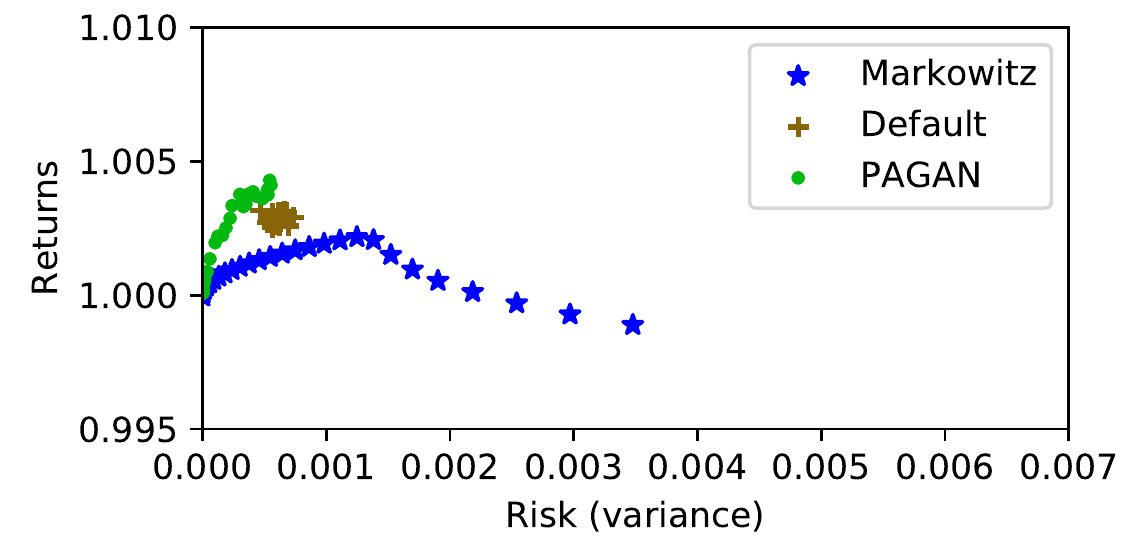}}
	}
	\subfloat[{\textit{usgen}, 20-days horizon.}]{
		\label{fig:resultUS}
		\resizebox{\sfigbig}{!}{\includegraphics{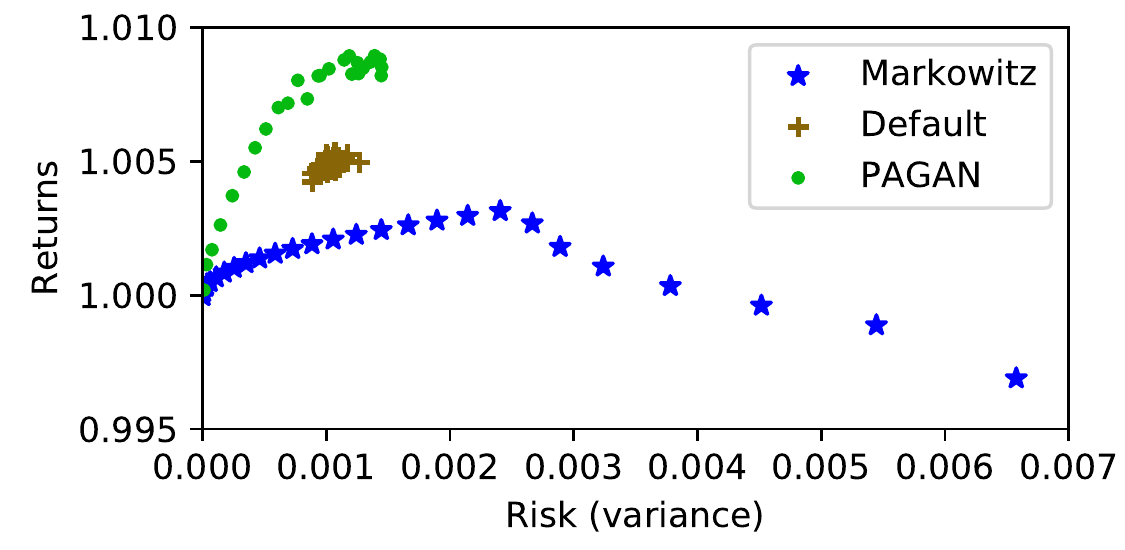}}
	}
	\qquad
	\subfloat[{\textit{eucar}, 5-days horizon.}]{
		\resizebox{\sfigbig}{!}{\includegraphics{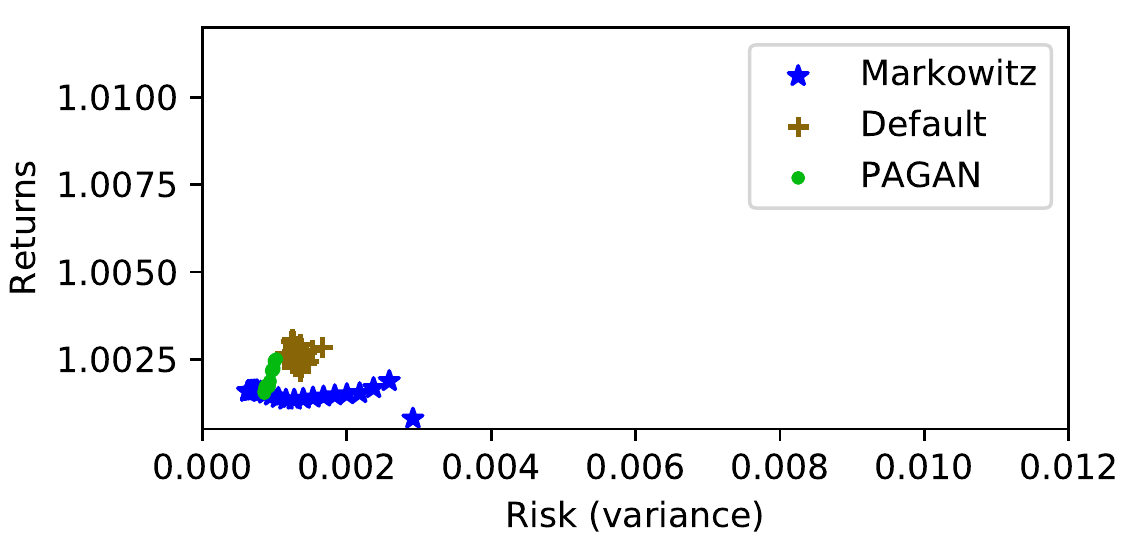}}
	}
	\subfloat[{\textit{eucar}, 10-days horizon.}]{
		\resizebox{\sfigbig}{!}{\includegraphics{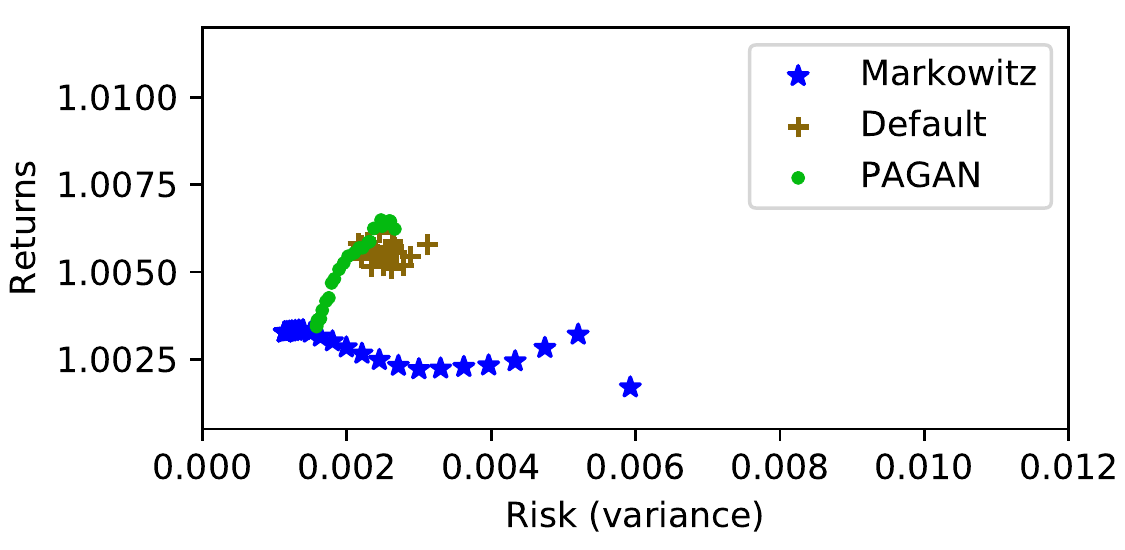}}
	}
	\subfloat[{\textit{eucar}, 20-days horizon.}]{
		\label{fig:resultEU}
		\resizebox{\sfigbig}{!}{\includegraphics{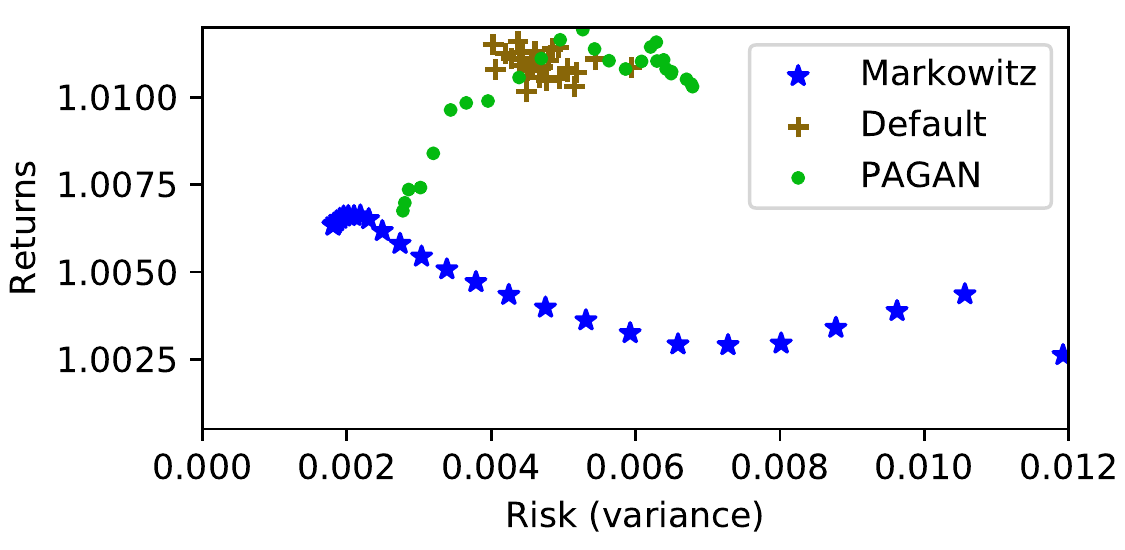}}
	}
	\caption{
		Returns-risk measured on the \textit{test} period by varying the risk levels (different points).
		The increasing-risk -- decreasing-returns behavior for \textit{Markowitz} happens because the returns probability distribution changes from the train to the test period
		thus models learnt on the train period may suggest to buy high-risk assets that loose value along the test period.
		PAGAN mitigates this issue by forecasting the future probability distribution $P(\forwardSequence | \backwardSequence)$ based on $\backwardSequence$.
	}
	\label{fig:result}
\end{figure*}

\begin{figure*}[t!]
	\centering
	\subfloat[{PAGAN \textit{usgen}.}]{
		\label{fig:diversificationPAGANUS}
		\resizebox{\sfigsmall}{!}{\includegraphics{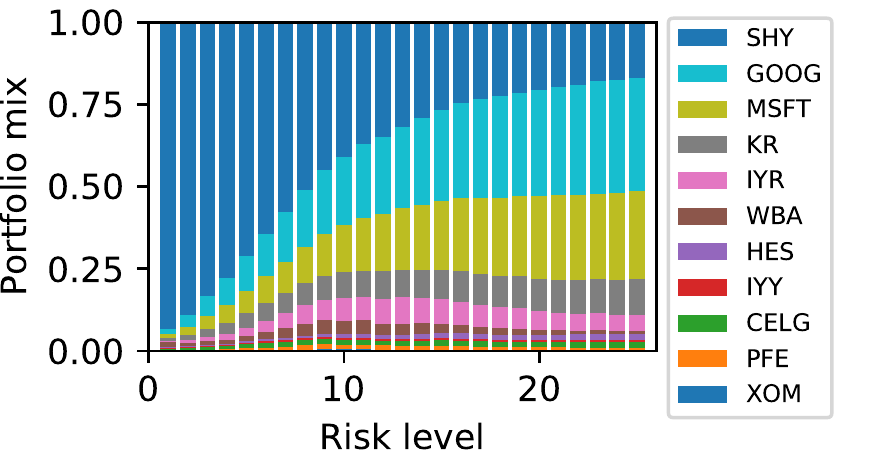}}
	}
	\subfloat[{Markowitz \textit{usgen}.}]{
		\label{fig:diversificationMUS}
		\resizebox{\sfigsmall}{!}{\includegraphics{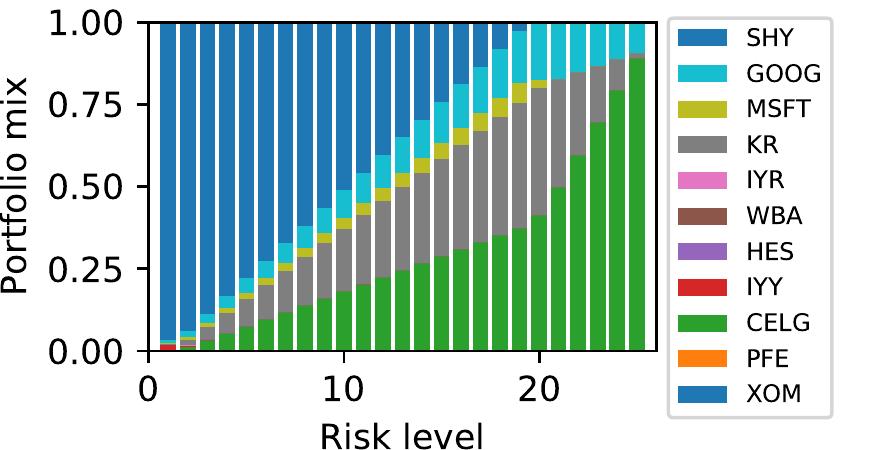}}
	}
	\subfloat[{PAGAN \textit{eucar}.}]{
		\label{fig:diversificationPAGANEU}
		\resizebox{\sfigsmall}{!}{\includegraphics{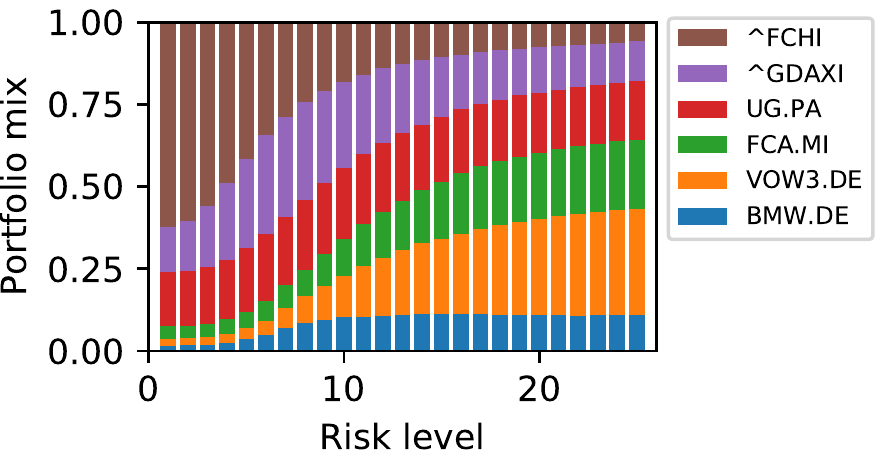}}
	}
	\subfloat[{Markowitz \textit{eucar}.}]{
		\label{fig:diversificationMEU}
		\resizebox{\sfigsmall}{!}{\includegraphics{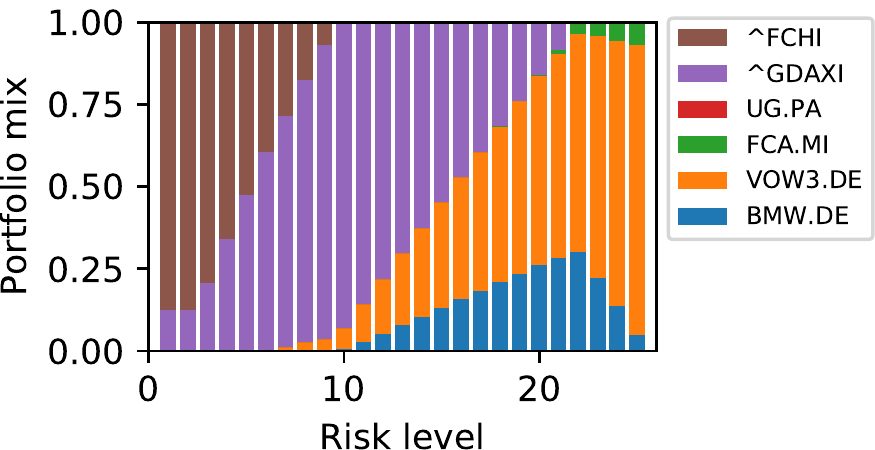}}
	}
	\caption{
		Diversification results for the considered portfolios by varying the target risk level $\riskLevel$ (x axis).
		These results are averaged along the whole test period. The portfolios from PAGAN are more diversified than those from Markowitz. PAGAN is able to systematically improve the returns achievable at a given risk.
	}
	\label{fig:diversification}
\end{figure*}

\begin{figure}[t!]
	\centering
	\subfloat[{PAGAN \textit{eucar}, September 2015.}]{
		\label{fig:diversificationPAGANEUsep}
		\resizebox{\sfigsmall}{!}{\includegraphics{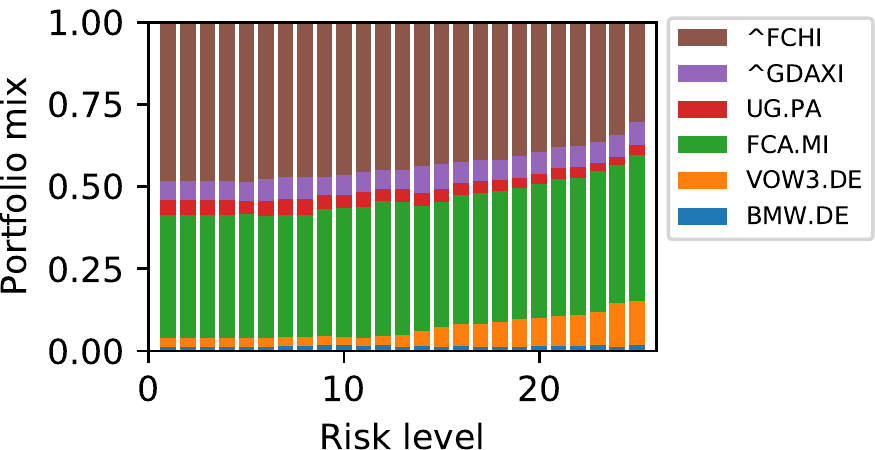}}
	}
	\subfloat[{PAGAN \textit{eucar}, October 2015.}]{
		\label{fig:diversificationPAGANEUoct}
		\resizebox{\sfigsmall}{!}{\includegraphics{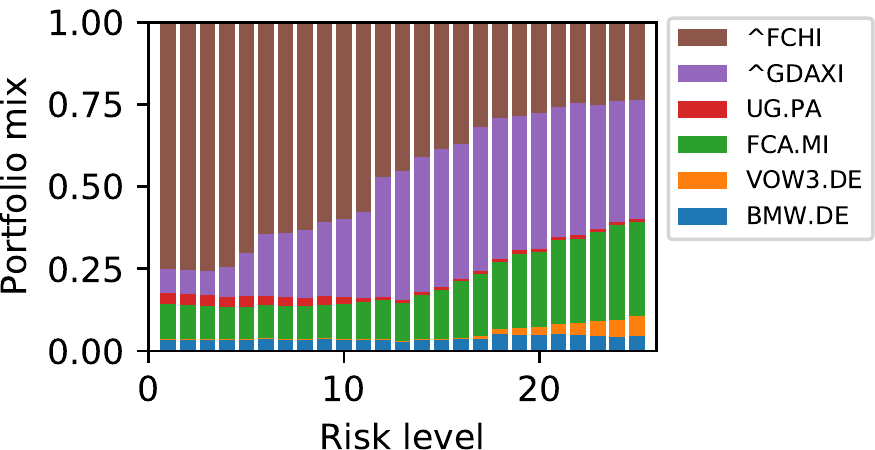}}
	}
	\caption{
		PAGAN diversification results before (September) and after (October) the Volkswagen emission scandal.
	}
	\label{fig:diversificationPAGANex}
\end{figure}

\subsection{Diversification results}

The output of the generative model are financial market simulations, such as the ones in Figure \ref{fig:resSimulations}.
Each simulation is carried out for all assets at once and represents a possible behavior of the marketplace. 
The goal is to draw possible scenarios and
to understand how the price of different assets are interacting, i.e. what may happen to one asset when another situation is presenting for another one.
PAGAN aims to capture non-linear dependencies between different trends and samples simulations
from the resulting multidimensional probability distribution.
For example, \textit{Sim\_4} in Figure \ref{fig:resSimulations} (thick dashed line)
shows a possible situation where both market indices (~$\hat{ }$~GDAXI, ~$\hat{ }$~FCHI)
have a sudden loss in the first half of April, followed by a quick re-bounce.
At the same time, UG.PA and FCA.MI accumulate significant losses by the end of the simulation (around the end of April). In this case (\textit{Sim\_4}) the best
would have been to buy VOW3.DE.
Yet, each simulation is just one possible realization of the probability distribution learnt by PAGAN. The goal of PAGAN is to enable us to investigate automatically several
different simulations in order to organize a portfolio diversification strategy. The solutions of the optimization problem defined in Equations \ref{eq:obj0}--\ref{eq:riskDefPAGAN}
generate a trade-off for the expected risk-return objective space given the probability distribution
modeled by a PAGAN-generated simulation set $\simulationSet$.
In this work, we set the number of simulations in $\simulationSet$ to 250.

We evaluate how good  the proposed approach is in diversifying the portfolio aiming at different time horizons $\windowForward$.
In particular we address horizons of one week (5 days),
two weeks (10 days), and 1 month (20 days). Figure \ref{fig:result}  shows the return-risk trade off achieved during the test period
by PAGAN and the reference benchmarks (Markowitz, default).

Since the market situation during the test period (starting in 2015-01) significantly diverges from the training period (ending in 2014-12),
the returns for Markowitz significantly decrease when aiming at high-risk solution. This happens because Markowitz assumes that the
future mean returns for a given asset equal the past ones (Assumption \ref{as:mark}). This approach let Markowitz buy risky assets that
were profitable along the train period and are losing in the test period. PAGAN does not assume the future probability distribution to be equal to the past one
because it learns to forecast the probability distribution given the most recent market situation. This significantly improves PAGAN results that are able to
generally provide higher returns when accepting a higher risk. Note that, given the very different market situation of the last few years, the default approach
of randomly buying assets provides not too bad solutions. Yet default is not capable of systematically trading off risk for returns leading to an unstructured cloud 
of solutions in the risk-return objective space. Default solutions are all drawn purely at random and their final results do not differentiate much from one another.

Figure \ref{fig:diversification} shows the diversifications proposed by PAGAN and Markowitz approaches for different risk levels given a one-month horizon $\windowForward=20$.
These diversifications are averaged along the whole test period.
PAGAN presents a smoother behavior. In fact, given the short-term conditioning window $\backwardSequence$,
PAGAN models the future probability distribution $P(\simulatedSequence | \backwardSequence)$ enabling it to quickly adapt
its diversifications to continuously and quickly changing market conditions.
As example, Figures \ref{fig:diversificationPAGANEUsep} and \ref{fig:diversificationPAGANEUoct} show
PAGAN diversifications averaged along September and October 2015. At the end of September 2015,
the Volkswagen emission scandal perturbs the European automotive market.
PAGAN already in September (Figure \ref{fig:diversificationPAGANEUsep}) allocates significantly fewer capital
to VOW3.DE than the overall average allocated along the whole test period (Figure \ref{fig:diversificationPAGANEU}).
In October, after the shock brought by the scandal to the automotive industries, PAGAN
shows a defensive strategy by allocating always more than 50\% of the overall capital to low volatility indices (~$\hat{ }$~FCHI, ~$\hat{ }$~GDAXI)
to avoid loosing capital on higher risk assets.


\begin{figure*}[t!]
	\centering
	\subfloat[{\textit{usgen}.}]{
		\label{fig:portfolioSimulationUS}
		\resizebox{\sfigwhole}{!}{\includegraphics{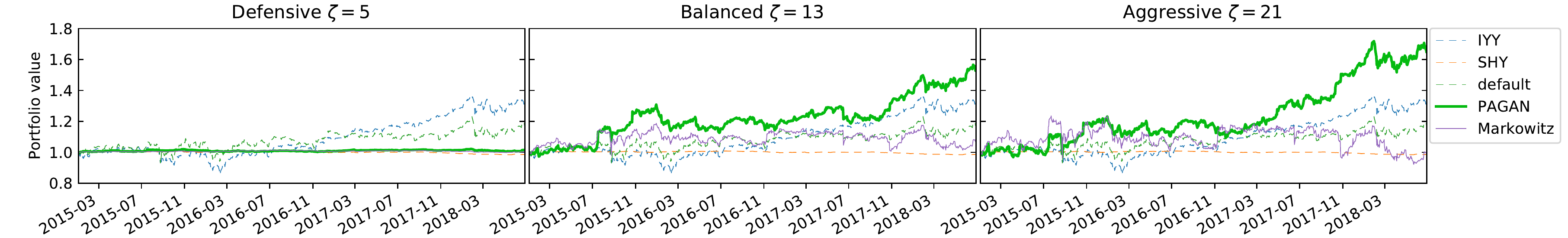}}
	}
	\qquad
	\subfloat[{\textit{eucar}.}]{
		\label{fig:portfolioSimulationEU}
		\resizebox{\sfigwhole}{!}{\includegraphics{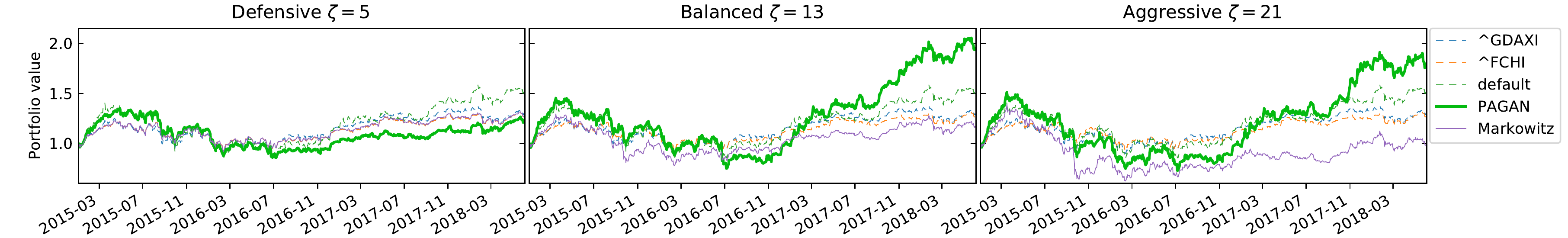}}
	}
	\caption{
		Portfolio values for different diversification risk settings (subplots).
		Reference benchmarks are shown with dashed lines, PAGAN and Markowitz approaches with solid lines. Note that in the defensive setting, the goal is not to maximize the portfolio returns but rather to reduce the risk (standard derivation of returns).
	}
	\label{fig:portfolioSimulation}
\end{figure*}

\begin{figure}[t!]
	\centering
	\subfloat[{\textit{usgen}.}]{
		\label{fig:portfolioAnalysisUS}
		\resizebox{\sfigbigbig}{!}{\includegraphics{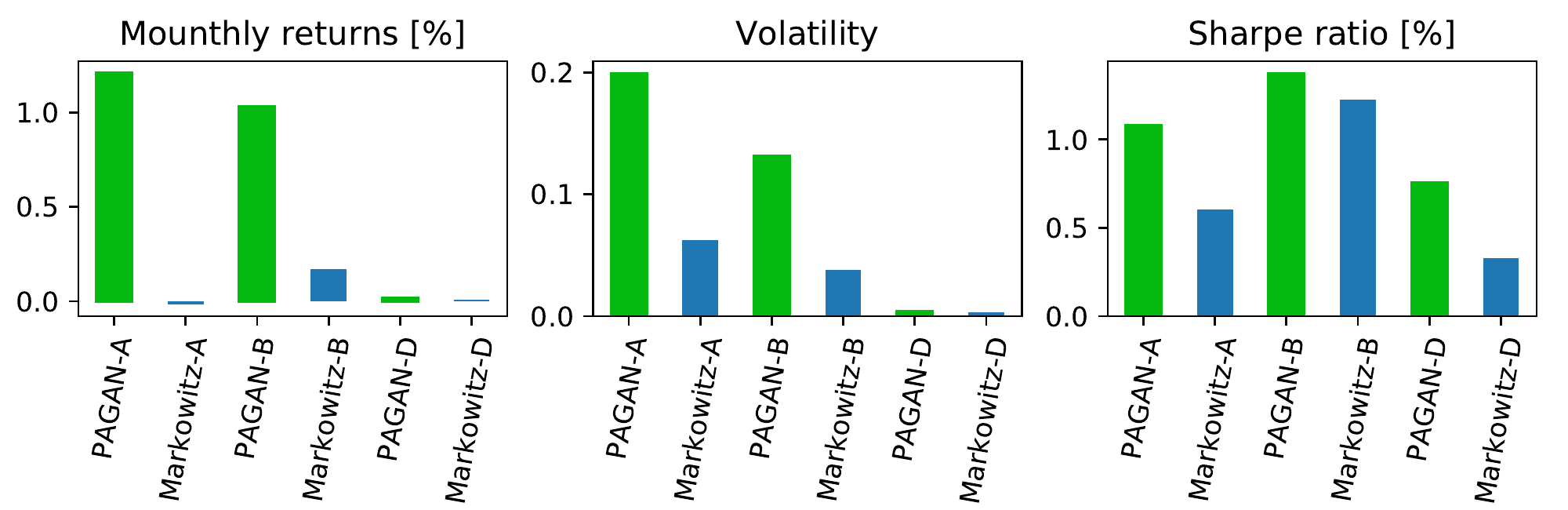}}
	}
	\qquad
	\subfloat[{\textit{eucar}.}]{
		\label{fig:portfolioAnalysisEU}
		\resizebox{\sfigbigbig}{!}{\includegraphics{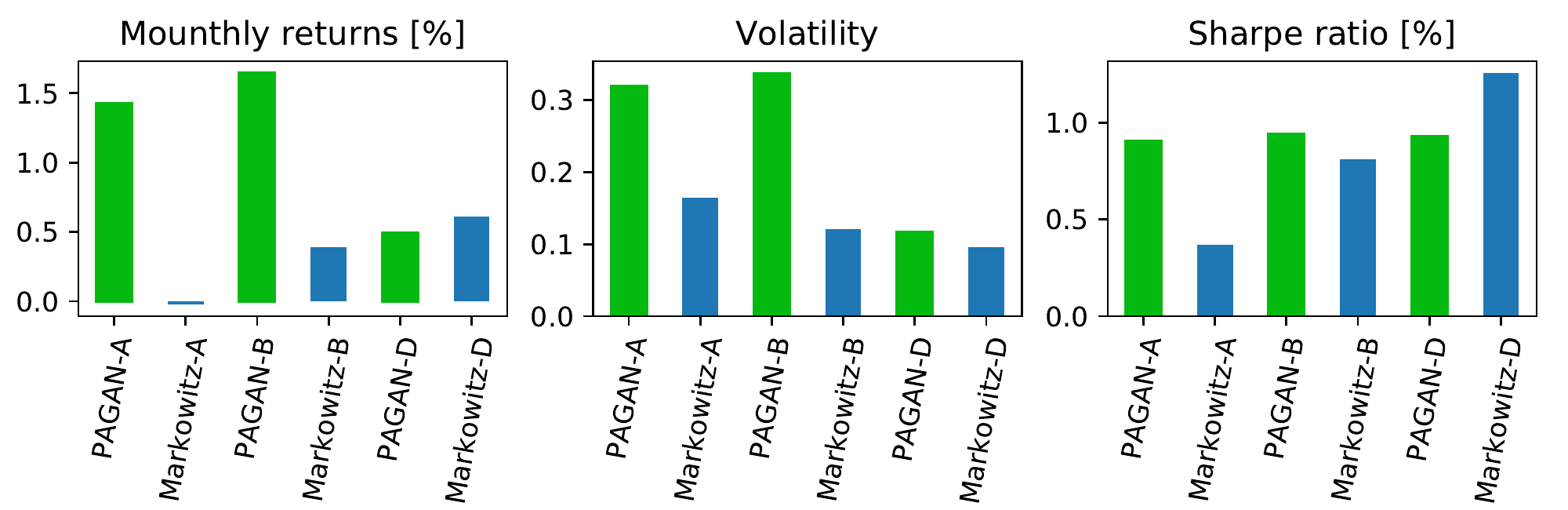}}
	}
	\caption{
		PAGAN surpasses the reference approach on financial-performance.
	}
	\label{fig:portfolioAnalysis}
\end{figure}

\subsection{Realized performance}

Let us analyse the financial performance of the considered optimization strategies by tracking the value of the corresponding portfolios along the test period.
We consider three settings for both Markowitz and PAGAN approaches.
A \textit{defensive} setting with risk level $\riskLevel=5$, a \textit{balanced} setting with $\riskLevel=13$, and an \textit{aggressive} setting with $\riskLevel=21$.
Over the considered $\totalRiskLevels=25$ risk levels, these settings account respectively for the 20\%, 50\%, and 80\% of the risk.
Since the performance for the default approach do not vary much by changing the risk level (Figure \ref{fig:result}), we only consider the \textit{balanced} version of the default approach.
We also include in the analysis portfolio-specific benchmarks such as \textit{a)} the Dow Jones Industrial Average ETF tracker (\textit{IYY}),
and the US Treasury Bond ETF tracker (\textit{SHY}) for \textit{usgen}, and \textit{b)} the German and French market indices (~$\hat{ }$~{GDAXI}, ~$\hat{ }$~{FCHI}) for \textit{eucar}.


We initialize each portfolio with an unitary value.
We consider an optimization horizon of 20 days to decide our diversification strategy, yet
we allow trading every day to keep the diversification to the suggested level. In fact,
a sudden variation of price for an asset changes the ratio of capital invested in this asset with respect to the others
if the number of shares is not adjusted accordingly.

Figure \ref{fig:portfolioSimulation} shows the simulation results along the test period.
PAGAN dominates the other approaches in terms of final portfolio value for the aggressive and balanced settings.
In the defensive setting, the goal is not to maximize the portfolio returns but rather to reduce the risk that is the
portfolio volatility (standard deviation of returns).
Figure \ref{fig:portfolioAnalysis} depicts the financial performance of PAGAN and Markowitz approaches including the different settings: aggressive (\textit{A}),
balanced (\textit{B}), and defensive (\textit{D}). PAGAN-A and PAGAN-B clearly dominate in terms of monthly returns.
This comes however at the cost of a higher risk (volatility). For a fairer comparison of the two approaches, we examine the annualized
sharpe ratio $\sharpe$ defined as
$\sharpe=(\return-\riskFreeReturn)/\sigma$, where $\return$ is the average annual return, $\riskFreeReturn$ is the risk-free
return\footnote{We consider as risk-free return a situation with no earning nor losses.}, and $\sigma$ is the volatility (standard deviation of annual returns).
In terms of Sharpe ratio (the higher, the better),
PAGAN surpasses the reference approach for all settings (A, B, D) for \textit{usgen} and for the A and B settings for \textit{eucar}.

\begin{figure}[t!]
	\centering
	\resizebox{\sfigbigbig}{!}{\includegraphics{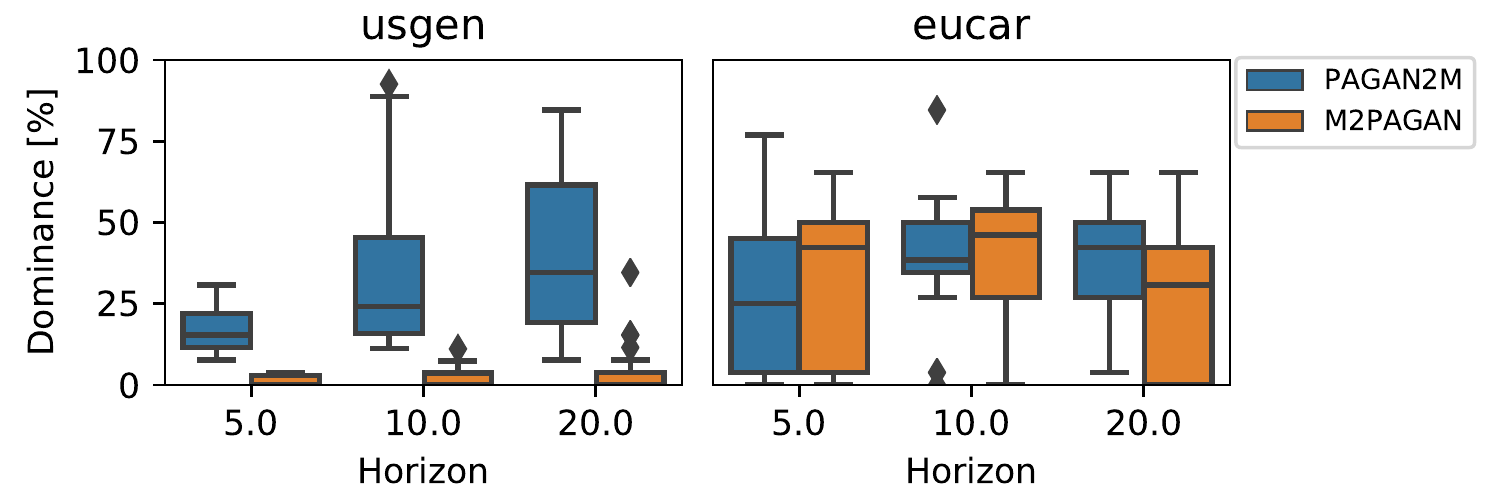}}
	\caption{{Repeatability: PAGAN (PAGAN2M) systematically surpasses Markowitz (M2PAGAN) for the \textit{usgen} portfolio. On the \textit{eucar} portfolio, \textit{PAGAN2M} and \textit{M2PAGAN} achieve comparable performance. }}
	\label{fig:repeatability}
\end{figure}

\subsection{Repeatability considerations}

GANs are known to be unstable, adversarial training often does not converge towards an equilibrium because of the non-linear dynamics
introduced by the differential equations implementing the learning algorithm \cite{mescheder2018}. This let the 
PAGAN model change during training. Furthermore, PAGAN results depend on the randomness introduced by the weight initialization. To analyze the repeatability
of the presented results we train several PAGAN models and compare them with Markowitz. This reference
approach is deterministic and, given the training data, it always produces the same results.
We proceed as follows.

Given a trained PAGAN model, we analyze the return--risk of PAGAN diversifications along the test period and we compare their results with the diversifications
suggested by Markowitz. Let $\sigma^{M}_{\riskLevel}$ and $\sigma^{P}_{\riskLevel}$ be the volatility observed for Markowitz and PAGAN diversifications
at risk level $\riskLevel$, and let $\return^{M}_{\riskLevel}$ and $\return^{P}_{\riskLevel}$ be the related returns.
We consider that PAGAN dominates Markowitz at risk level $\riskLevel$ if PAGAN results are better for at least one of the two metrics and not worse for the other:
$\big(\sigma^{P}_{\riskLevel}\leq\sigma^{M}_{\riskLevel} \bigwedge \return^{P}_{\riskLevel}>\return^{M}_{\riskLevel}\big) \bigvee \big(\sigma^{P}_{\riskLevel}<\sigma^{M}_{\riskLevel} \bigwedge \return^{P}_{\riskLevel}\geq\return^{M}_{\riskLevel}\big)$. For every trained PAGAN model, we compute the percentage of risk levels $\riskLevel$ for which PAGAN dominates Markowitz (\textit{PAGAN2M}),
and the opposite (\textit{M2PAGAN}). Figure \ref{fig:repeatability} shows the distribution of these metrics in box-plot for the two portfolios when considering different optimization horizons.
In general, \textit{PAGAN2M} increases for higher horizon demonstrating the difficulties of Markowitz in modeling long-term situations and the advantage of using the proposed PAGAN model.
PAGAN systematically surpasses Markowitz for the \textit{usgen} portfolio with values of \textit{PAGAN2M} much higher than \textit{M2PAGAN}. On the \textit{eucar}
portfolio, \textit{PAGAN2M} and \textit{M2PAGAN} are partially overlapping. For \textit{eucar}, Markowitz achieves sometimes performance comparable to PAGAN. Yet, when aiming at an horizon of 20 days,
PAGAN in average outperforms Markowitz. For a fair comparison, all results presented in this paper along the previous sections are gathered
starting from the PAGAN model that achieved the median result for the \textit{PAGAN2M} metric for an horizon of 20 days. Additional results can be found in Appendix \ref{sec:addRes}.



\section{Conclusion}
In this work we presented a pioneering study about portfolio analysis with generative adversarial networks (PAGAN).
A key novelty is that the proposed approach is explicitly intended to address the increasing challenges under high efficient market, i.e. when assuming that all information available today are already represented in current asset prices leaving limited room to medium--long term price trend forecast. 
We introduced the Markowitz framework to demonstrate how, under these conditions,
it is still possible to take an educated guess on future returns probability distribution and use this guess to
decide about a diversification strategy to optimize a portfolio for minimizing its risk and maximizing its expected returns.
By relying on these considerations we proposed to apply a generative network to learn modeling
the market uncertainty in its complex multidimensional form, such to let the deep-learning system embed
non-linear interactions between different assets. The results demonstrate a clear advantage with respect
to the state of the art in portfolio optimization theory. In particular, the proposed approach is able
to expose to the final user the possibility of selecting a target risk level and to suggest a specific diversification
given the current market situation. Compared to the Markowitz modern portfolio optimization approach, we systematically
achieve better performance in terms of both, expected return maximization, and risk minimization.

\bibliographystyle{unsrt}
\bibliography{biblio}

\appendix

\section{PAGAN hyperparameter settings}
\label{sec:hyperparams}
In this section we discuss the details of PAGAN networks and optimizer parameters.
Detailed architectural parameters for the generator $\generator$, and the discriminator $\discriminator$ are
listed in Tables \ref{tab:parametersGen}, and \ref{tab:parametersDis}.

\textbf{Optimizer.}  PAGAN generator and discriminator (Figure \ref{fig:architecture}) are trained with Adam's optimizer with learning rate $2\times10^{-5}$, and $\beta_1 = 0.5$.
PAGAN models have been trained for 15'000 epochs.

\begin{table}[b!]
	\centering
	\scriptsize
	\caption{List of generator parameters.
	}
	\setlength\tabcolsep{0.15cm}
	\begin{tabular}{|c||   c| }
		\hline
		\textbf{Description} & \textbf{Value} \\
		\hline
		\hline
		Size of latent vector $\latent$ & $2 \times \assets$ \\
		\hline
		Analysis window $\window=\windowBackward+\windowForward$ & $60$ \\
		\hline
		Forward window $\windowForward$ & $20$ \\
		\hline
		Backward window $\windowBackward$ & $40$ \\
		\hline
		Convolution (conv.) layers in \textit{conditioning} & $4$ \\
		\hline
		Input and output channels in all conv. layers in \textit{conditioning} & $2 \times \assets$ \\
		\hline
		Dense layer output size in  \textit{conditioning} & $\assets$ \\
		\hline
		\hline
		Dense layer output size in  \textit{simulator} & $ \windowForward \times \assets $ \\
		\hline
		Transpose conv. layers in  \textit{simulator} & $2$ \\
		\hline
		Input channels in the first transpose conv. layers in  \textit{simulator} & $4 \times \assets$ \\
		\hline
		Output channels in the first transpose conv. layers in  \textit{simulator} & $2 \times \assets$ \\
		\hline
		Input channels in the second transpose conv. layers in  \textit{simulator} & $2 \times \assets$ \\
		\hline
		Output channels in the second transpose conv. layers in  \textit{simulator} & $ \assets$ \\
		\hline
		\hline
		Layers' activation function in \textit{conditioning} and \textit{simulator} & \textit{ReLu} \\
		\hline
		Conv. (and transpose conv.) kernel length in \textit{conditioning} and \textit{simulator} & $5$ \\
		\hline
		Conv. (and transpose conv.) stride in \textit{conditioning} and \textit{simulator} & $2$ \\
		\hline
	\end{tabular}
	\label{tab:parametersGen}
\end{table}

\textbf{Generator.}
The \textit{conditioning} network in the generator $\generator$ (Figure \ref{fig:genArch}) is composed
of 4 consecutive convolutional layers with a constant number of channels. We use a convolution stride of two
to iteratively compresses the inner representation of the backward sequence $\backwardSequence$ by
halving the time resolution at each layer. In generative models the compression of the resolution by means of strided convolution
is very common \cite{radford2015}.
The output of these convolutional layers is then processed by means of a dense layer,
and then concatenated with the analysis vector $\analysisVector$ and the latent vector $\latent$.

The \textit{simulator} network (Figure \ref{fig:genArch}) includes a first dense layer followed by a sequence of transpose convolution layers.
As in traditional GANs \cite{radford2015}, in PAGAN the \textit{generator} applies transpose convolutions with strides of two to iteratively compress the number of channels
and expand the time resolution. We apply two transpose-convolution layers. The output of the last transpose convolution 
has a number of channels equal to the number $\assets$ of assets in the target portfolio,
and a time resolution equal to the simulation horizon $\windowForward$. At every layer of the generative transpose convolutions
we halve the number of channels and double the time resolution. Thus the output of the dense layer right before the first transpose convolution has by construction
$2^{\simTconvLayers} \times \windowForward/2^{\simTconvLayers} \times \assets = \windowForward\times\assets$ outputs, where $\simTconvLayers$ is the number
of transpose-convolution layers.

\textbf{Discriminator.}
The discriminator (Figure \ref{fig:disArch}) is implemented with a traditional sequence of convolutional layers that iteratively halves the time resolution
and doubles the channels \cite{radford2015}. The last dense layer has a single output,
a critic value as proposed in WGAN-GP \cite{gulrajani2017}.
We apply spectral normalization \cite{miyato2018} on the kernel weights of
the discriminator since this procedure improves the convergence and stability of GANs \cite{kurach2018,brock2018}.

\begin{table}[b!]
	\centering
	\scriptsize
	\caption{List of discriminator parameters.
	}
	\setlength\tabcolsep{0.15cm}
	\begin{tabular}{|c||   c| }
		\hline
		\textbf{Description} & \textbf{Value} \\
		\hline
		\hline
		Analysis window $\window$ & $60$ \\
		\hline
		Convolution layers in $\discriminator$ & $5$ \\
		\hline
		Input channels for the $i$th convolution layer in $\discriminator$ & $\assets \times 2^{i-1}$ \\
		\hline
		Output channels for the $i$th convolution layer in $\discriminator$ & $\assets \times 2^{i}$ \\
		\hline
		Layers' activation function in $\discriminator$ & \textit{LeakyReLu} \\
		\hline
		Convolution kernel length & $5$ \\
		\hline
		Convolution stride & $2$ \\
		\hline
	\end{tabular}
	\label{tab:parametersDis}
\end{table}


\section{Additional results}
\label{sec:addRes}

Adversarial training defines a non-linear dynamic system where generator continuously adapts to the
discriminator and vice versa. This leads to a series of convergence problems and results may change
depending on the random initialization. For this reason we have retrained
several PAGAN models to make sure that results are meaningful.
The results presented in Section \ref{sec:res} refer to the PAGAN models
that produced the median value of the PAGAN2M metric (Figure \ref{fig:repeatability}),
thus they are fair and representative.
We here introduce some additional results obtained from other random initializations and the same experimental settings.

\begin{figure*}[t!]
	\centering
	\subfloat[{\textit{usgen}, 80th percentile of PAGAN2M.}]{
		\label{fig:resUS80}
		\resizebox{\sfigbig}{!}{\includegraphics{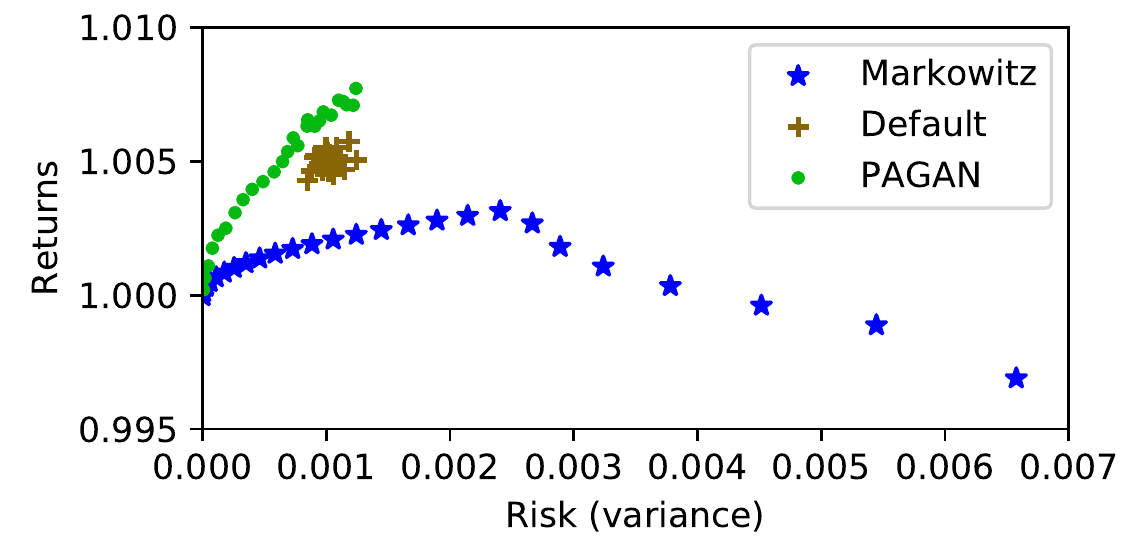}}
	}
	\subfloat[{\textit{usgen}, 50th percentile of PAGAN2M.}]{
	\resizebox{\sfigbig}{!}{\includegraphics{./results/tradeoff-us-20.pdf}}
	}
	\subfloat[{\textit{usgen}, 20th percentile of PAGAN2M.}]{
		\label{fig:resUS20}
		\resizebox{\sfigbig}{!}{\includegraphics{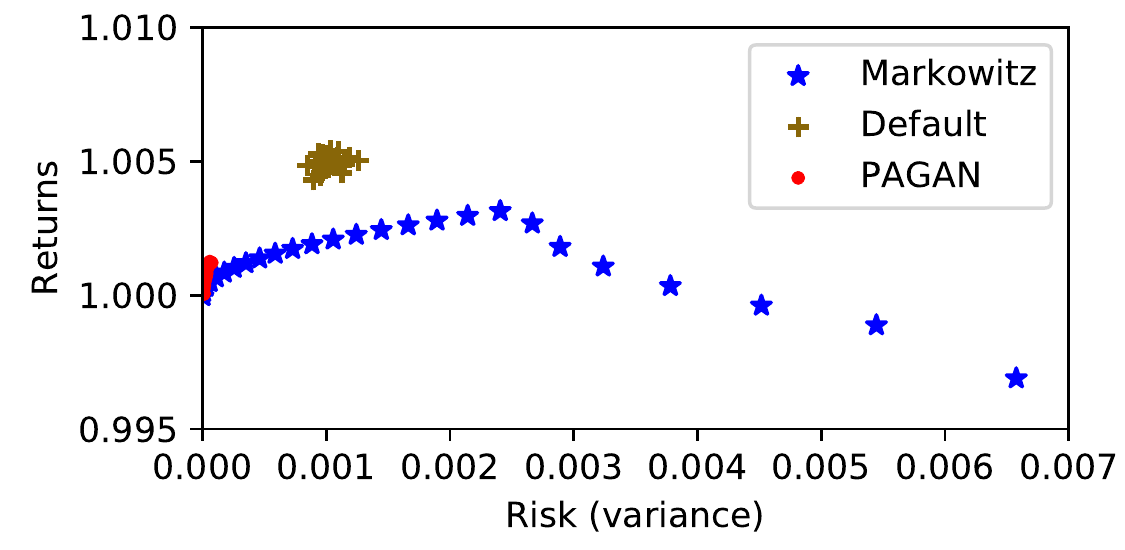}}
	}
\qquad
	\subfloat[{\textit{eucar}, 80th percentile of PAGAN2M.}]{
		\resizebox{\sfigbig}{!}{\includegraphics{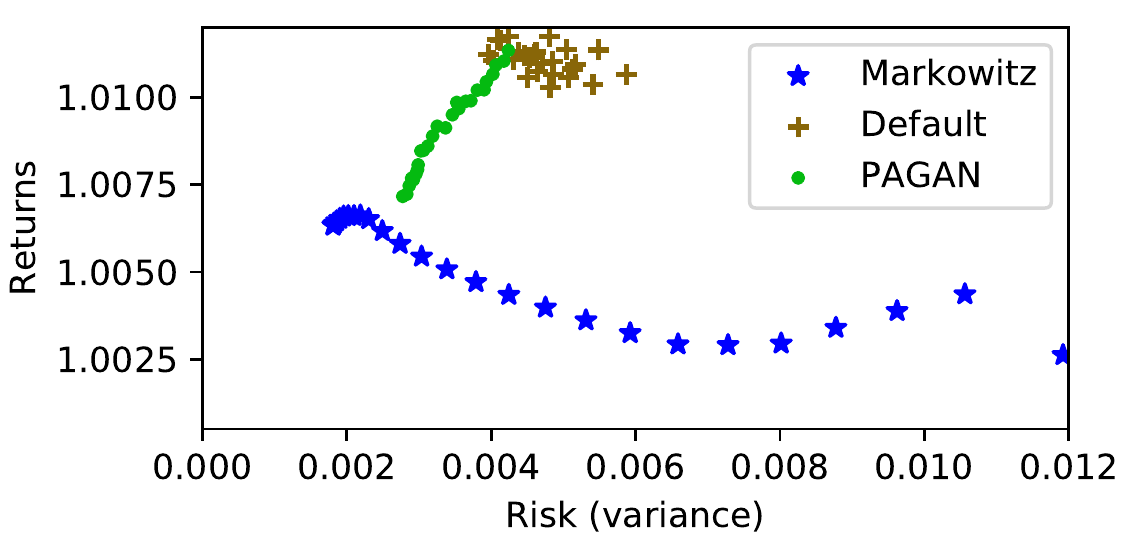}}
	}
	\subfloat[{\textit{eucar}, 50th percentile of PAGAN2M.}]{
	\resizebox{\sfigbig}{!}{\includegraphics{./results/tradeoff-eu-20.pdf}}
	}
	\subfloat[{\textit{eucar}, 20th percentile of PAGAN2M.}]{
	\resizebox{\sfigbig}{!}{\includegraphics{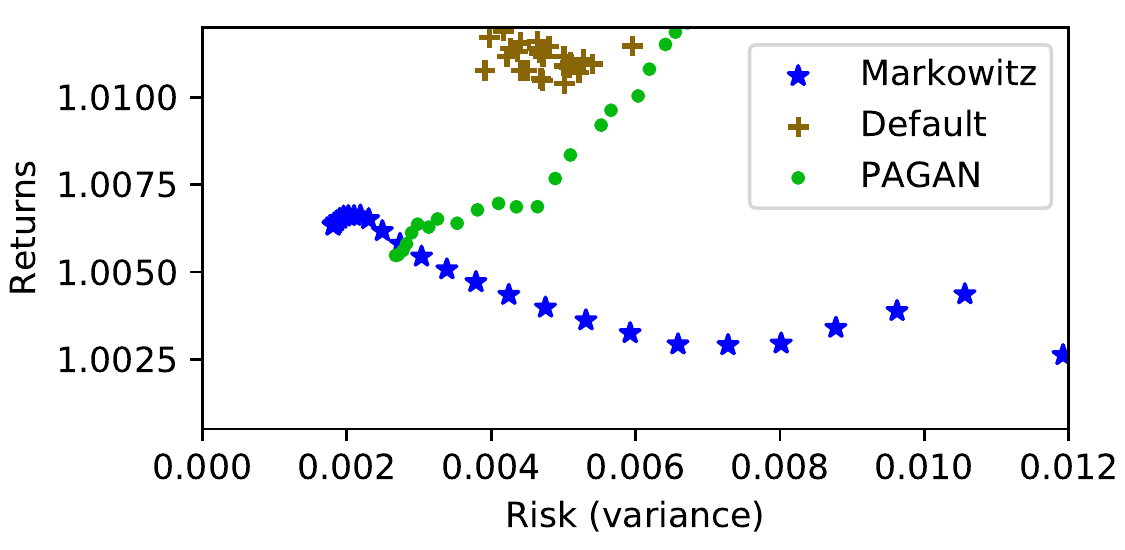}}
	}
	\caption{
		Returns--risk trade of measured on the \textit{test} period by varying the risk levels (different points) considering a 20-days optimization horizon.
		PAGAN models generating these results correspond to the 80th, 50th, and 20th percentiles of the PAGAN2M metric (Figure \ref{fig:repeatability}).
	}
	\label{fig:resPercentiles}
\end{figure*}


\begin{figure*}[t!]
	\centering
	\subfloat[{80th percentile of PAGAN2M.}]{
		\label{fig:resAddSimulations80}
		\resizebox{\sfigwhole}{!}{\includegraphics{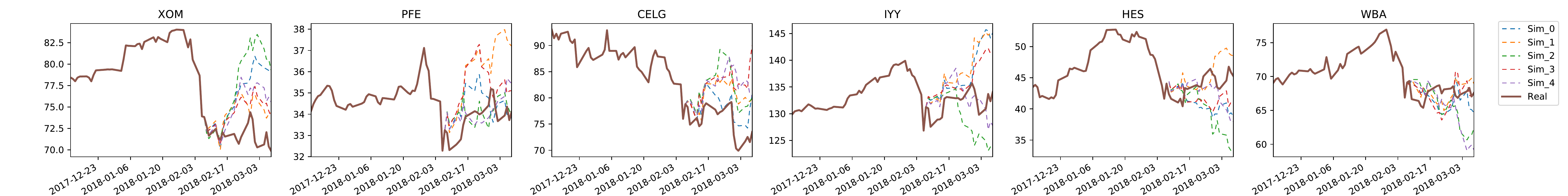}}
	}
	\qquad
	\subfloat[{50th percentile of PAGAN2M.}]{
		\label{fig:resAddSimulations50}
		\resizebox{\sfigwhole}{!}{\includegraphics{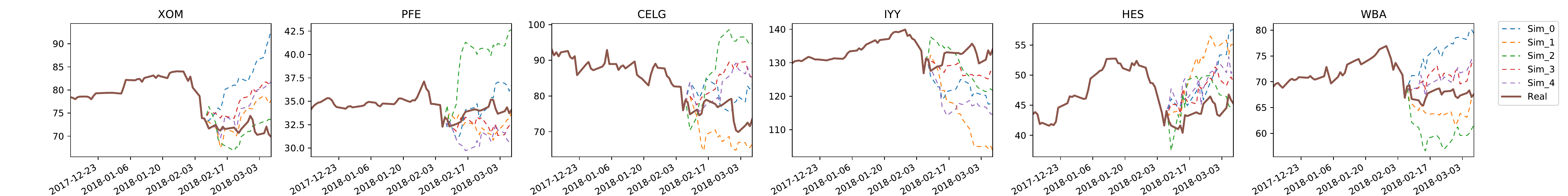}}
	}
	\qquad
	\subfloat[{20th percentile of PAGAN2M.}]{
	\label{fig:resAddSimulations20}
	\resizebox{\sfigwhole}{!}{\includegraphics{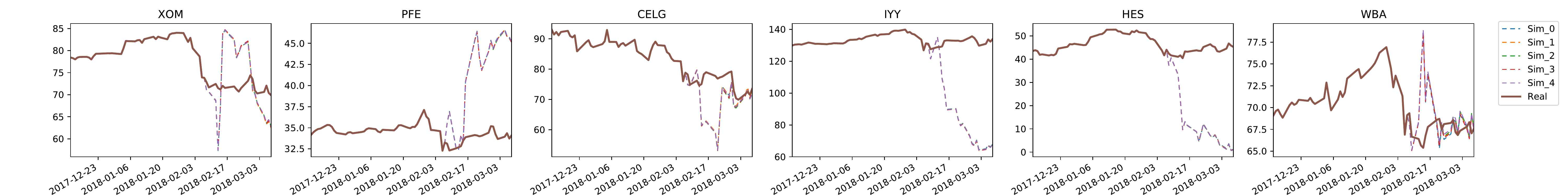}}
	}
	\caption{
		Representative simulations for six assets in the \textit{usgen} portfolio when using different PAGAN models.
		PAGAN model at the 20th percentile of the PAGAN2M metric mode-collapsed
		and keep on repeating the same simulation (Figure \ref{fig:resAddSimulations20}).
	}
	\label{fig:resAddSimulations}
\end{figure*}

Figure \ref{fig:resAddSimulations} shows the risk-returns trade off for the PAGAN models returning the
80th, 50th, and 20th percentiles of the PAGAN2M metric for the two portfolios (the 50th percentile is the median as in Figures
\ref{fig:resultUS}, \ref{fig:resultEU}).
PAGAN's results at the 80th percentile of the PAGAN2M metric are fairly similar to the ones at the 50th percentile.
Yet, not all PAGAN models
return good results, for example Figure \ref{fig:resUS20} shows a situation where PAGAN is not able
to provide good solutions and only suggests low-risk diversifications without exposing a good risk--returns trade of
(PAGAN solutions have been highlighted in red). In this case, the problem is strictly related
to the adversarial training, the PAGAN generator has mode collapsed and keep on repeating the same simulation
$\simulatedSequence$ that is not representative of the actual probability
distribution of $\forwardSequence$. Figure \ref{fig:resAddSimulations},
depicts example simulations obtained for the PAGAN models related with the results
in Figures \ref{fig:resAddSimulations}.

Mode collapse happens rarely with our experimental
settings (Appendix \ref{sec:hyperparams}) and only in the lower tail of
the PAGAN2M metric (Figure \ref{fig:repeatability}),
i.e. for the least performing PAGAN models.
Mode collapse can be easily identified, e.g. with a graphical inspection of the simulations, Figure \ref{fig:resAddSimulations20}.
Removing the collapsed models from our analysis would further improve the results in favour of the PAGAN approach.
Yet, for a fair evaluation, in this work all the PAGAN models including the collapsed ones
are considered
and results are captured in the plots of Figure \ref{fig:repeatability}.
Mode collapse is a well known problem in GANs \cite{miyato2018,Goodfellow2014}.
To the best of our knowledge, there is not yet a generally accepted and well established solution.


\end{document}